# Searching for quantum non-thermodynamic phenomena

## Yu Qiao[1,2,*]

[1] *Program of Materials Science and Engineering, University of California – San Diego, La Jolla, CA 92093, U.S.A.*

[2] *Department of Structural Engineering, University of California – San Diego, La Jolla, CA 92093-0085, U.S.A.*

[*] *Email: yqiao@ucsd.edu*

**Abstract.** Recent research on the fundamentals of statistical mechanics has led to an interesting discovery [1-3]: With locally nonchaotic barriers, as Boltzmann's H-theorem is inapplicable, there exist nontrivial non-thermodynamic systems that can produce useful work by absorbing heat from a single thermal reservoir without any other effect, thereby breaking the boundaries of the second law of thermodynamics. The previous analyses used classical mechanical models. In the current investigation, the study is extended to quantum mechanics. First, we reiterate that the Fermi-Dirac distribution and the Bose-Einstein distribution are compatible with the generalized Maxwell's relations, which demonstrates the general robustness of the framework of quantum statistical mechanics. Next, we analyze a set of simple-step scattering problems. When the system is in contact with a thermal reservoir, a bound state inherently follows the second law of thermodynamics, while a scattering state may not. The root cause is associated with the nonlocal nature of the wave function. It implies that the non-thermodynamic phenomena favor unquantized energy and localized wave packets, exhibiting a tendency to occur in "semiclassical" setups.

## 1. Introduction

### 1.1 Background: nonchaoticity and thermodynamic limit

Recent research on the fundamentals of statistical mechanics [1-3] revealed that with locally nonchaotic components, certain macroscopic systems may spontaneously reach nonequilibrium steady states and have unusual energy properties. They can produce useful work by absorbing heat from a single thermal reservoir with no other effect, thereby breaking the boundaries of the second law of thermodynamics.

Traditionally, the second law of thermodynamics was "derived" from Boltzmann's H-theorem [4]. The H-theorem attempts to explain how, in an isolated system wherein all the governing equations are time-reversible, the increase of entropy is irreversible. The key concept is the hypothesis of molecular chaos [5]. It assumes that upon random particle-particle collisions, the degree of correlation of particle trajectories varies, so that the probability of the evolution paths of system states is time-asymmetric. However, without the widespread particle-particle collisions, this framework is not applicable.

In fact, in classical mechanics, it is well known that thermodynamics has its limits and cannot be applied in all the scenarios [6], especially for some nonchaotic particle movements. One example is a single elastic particle moving in a box [2]. The box is formed by thermal walls, immersed in a thermal reservoir at a constant temperature $T$. The system is ergodic. It cannot relax to thermal equilibrium, in the sense that at the steady state, the time-average particle kinetic energy is lower than that of the environment. The probability of finding the particle at a speed $v$ ($\bar{p}_v$) is proportional not only to the Maxwell-Boltzmann distribution $p_{\mathrm{MB}}(v)$ but also to $1/v$, i.e., $\bar{p}_v \propto p_{\mathrm{MB}}/v$. At a high speed, the particle tends to rapidly collide with the box wall and release heat to the environment; at a low speed, the particle tends to stay long in the interior. Hence, compared to the equilibrium state, it is more likely to find the particle at a lower $v$. Quite often, such phenomena are regarded as "trivial", because the systems are small and no useful work can be produced.

## 1.2 Recent research on classical mechanical non-thermodynamic models

Figure 1(a,b) shows a large-sized ideal-gas model [1,7]. The system is closed and immersed in a thermal reservoir at $T$. In a uniform gravitational field ($g$), across a low-height step, a large number of elastic particles randomly move on two horizontal shelves. The step height ($z_s$) is much less than the nominal particle mean free path ($\lambda_F$). In the step, particle-particle collisions are sparse, so that the particle trajectories tend to be independent of each other, i.e., the step is a locally nonchaotic energy barrier. As a result, at the steady state, the ratio of particle number density between the upper and lower shelves ($\bar{\rho}$) is not the Boltzmann factor ($\delta_0 = e^{-\beta m g z_s}$), but rather $\delta_1 = 1 - \mathrm{erf}(\sqrt{\beta m g z_s})$, where $\beta = 1/(k_B T)$, $k_B$ is the Boltzmann constant, and $m$ is the particle

mass. The lower-shelf area ($A_\text{P}$) can be compressed or expanded by the in-plane pressure ($P$), and the upper shelf can be raised or lowered by the support force ($F_\text{G}$). An isothermal cycle may be designed to alternately adjust $A_\text{P}$ and $z_\text{s}$. Only when the system is in equilibrium ($\bar{\rho} = \delta_0$), can the total produced work ($W_\text{P}$) be equal to the total consumed work ($W_\text{G}$). Since $\bar{\rho} \to \delta_1$, $W_\text{P} > W_\text{G}$. The system cyclically absorbs heat from the environment (a single thermal reservoir) and converts it to useful work.

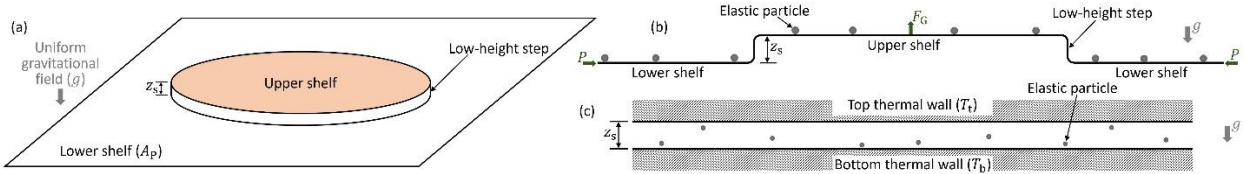

**Figure 1.** Two classical mechanical non-thermodynamic models. They are unrelated to Maxwell's demon or Feynman's ratchet. **(a)** Elevated view and **(b)** side view of the two-shelf ideal-gas model [1,7]. A large number of elastic particles (not shown in the elevated view) randomly move in a uniform gravitational field ($g$) on two large horizontal shelves, across the low-height step. The step height ($z_\text{s}$) is much less than the nominal particle mean free path ($\lambda_\text{F}$). The steady-state ratio of particle number density between the upper and lower shelves ($\bar{\rho}$) is intrinsically non-Boltzmannian. **(c)** A large number of elastic particles randomly move in a uniform gravitational field ($g$) between two horizontal thermal walls [2]. The gap thickness ($z_\text{s}$) is much less than the nominal $\lambda_\text{F}$. As particle-particle collisions are sparse in the gap, the particle trajectories tend to be independent of each other. At the steady state, even when the top-wall temperature ($T_\text{t}$) is equal to the bottom-wall temperature ($T_\text{b}$), the average particle kinetic energy is a function of height, i.e., the gas phase cannot relax to thermal equilibrium.

The steady state in Figure 1(a,b) is *intrinsically* nonequilibrium ($\bar{\rho} \to \delta_1$), fundamentally distinct from the conventional nonequilibrium processes [6]. There is no external driving force that drives the system away from equilibrium, e.g., a temperature or pressure gradient maintained by the environment. The second law of thermodynamics forbids such a system from being in nonequilibrium: nonequilibrium and lack of energetic penalty (e.g., being isolated or isothermal) should be mutually exclusive. In essence, as particle-particle collisions rarely happen in the narrow energy barrier, the system does not have a mechanism to approach thermodynamic equilibrium (i.e., $\bar{\rho} \neq \delta_0$), and entropy ($S$) cannot increase to the equilibrium maximum ($S_\text{eq}$).

Figure 1(c) depicts another example [2]. In a uniform gravitational field ($g$), a large number of elastic particles randomly move in between two large horizontal thermal walls. The top-wall temperature and the bottom-wall temperature are denoted by $T_\text{t}$ and $T_\text{b}$, respectively. The gap thickness $z_\text{s} \ll \lambda_\text{F}$. As particle-particle collisions are negligible in the gap, the particles tend to ascend or descend individually, and the system state is dominated by the particle-wall collisions.

At the steady state, even when $T_t = T_b$, the average particle kinetic energy is a function of height; that is, the gas phase cannot relax to thermal equilibrium. A remarkable consequence is that, if $T_t$ is mildly higher than $T_b$, heat can be spontaneously and continuously transported by the gas particles from the cold side (the bottom wall) to the hot side (the top wall), contradicting the refrigeration statement of the second law of thermodynamics.

The two models in Figure 1 are not variants of the traditional "counterexamples" of the second law of thermodynamics. The traditional setups can be represented by either Maxwell's demon [8] or Feynman's ratchet [9]. They may appear to be incompatible with statistical mechanics yet actually agree with it. Maxwell's demon is nonequilibrium, but nonspontaneous; Feynman's ratchet is spontaneous, but not nonequilibrium. On the contrary, Figure 1(a,b) and Figure 1(c) are both spontaneous and intrinsically nonequilibrium. The essential factor is nonchaoticity. Without chaotic particle-particle interactions to induce a long-range correlation of particle states, the local characteristics of kinetics becomes important.

To achieve a significant nonequilibrium effect, Figure 1(a,b) and Figure 1(c) demand an ultrahigh $g$. For example, if the gas particles are ambient air molecules, $g$ needs to be more than $10^{11}$ m/s$^2$, at the level of neutron stars. While it is difficult to directly observe the out-of-equilibrium behavior in gravity, this hurdle may be circumvented by changing the weak gravitational force to a stronger thermodynamic force, such as the Columb force. Another method is to use time-dependent locally nonchaotic entropy barriers, as demonstrated by the experiments on a nanoporous polyamide membrane one-sidedly surface-treated with bendable organic chains [3] and the temperature variation associated with the movement of a nonwetting liquid in the nanoporous silica particles [2].

## 1.3 Consistency with the principle of maximum entropy

Any model that does not follow the second law of thermodynamics must answer the following question: without an energetic penalty, how can entropy not reach its maximum possible value? Entropy measures probability. The principle of maximum entropy represents the basic logic that the most probable system state has the highest probability to occur, which should be considered "unconditionally" true.

For an isolated chaotic system, assume that at the $i$-th energy level $\epsilon_i$ ($i = 1,2,3 \ldots$), the number of particles ($n_i$) is much less than the density of states ($g_i$). In classical mechanics, the number of possible microstates is [5]

$$\Omega \approx \prod_i \frac{g_i^{n_i}}{n_i!} \tag{1}$$

Based on Stirling's formula [10], for large $n_i$, the Gibbs entropy is

$$S = k_B \ln \Omega \approx k_B \sum_i n_i \ln \frac{g_i}{n_i} \tag{2}$$

With extensive particle-particle collisions, $n_i$ is not explicitly dependent on kinetic theory and is subject to only two general constraints:

$$\sum_i n_i = N_{\text{tot}} \text{ and } \sum_i n_i \epsilon_i = U, \tag{3}$$

where $N_{\text{tot}}$ is the total particle number and $U$ is the internal energy. The second law of thermodynamics dictates that $S$ must be maximized, i.e.,

$$\frac{\partial \mathcal{L}}{\partial n_i} = 0 \tag{4}$$

where $\mathcal{L} = k_B \ln \Omega - \alpha_n (\sum_i n_i - N_{\text{tot}}) - \beta_n (\sum_i n_i \epsilon_i - U)$ is the Lagrangian, and $\alpha_n$ and $\beta_n$ are the Lagrange multipliers. The solution of Equation (4) is the Maxwell-Boltzmann distribution*

$$\frac{n_i}{g_i} = \varphi_m e^{-\beta \epsilon_i} \tag{5}$$

where $\varphi_m = e^{-(1+\alpha_n/k_B)}$, and $\beta = \beta_n/k_B = 1/(k_B T)$. Substituting Equation (5) into Equation (2) gives the maximum possible entropy that the system can ever reach:

$$S_{\text{eq}} = k_B N_{\text{tot}} \ln \varphi_m^{-1} + k_B \beta U \tag{6}$$

Any other distribution of $n_i$ (e.g., Equation 9 below) would lead to a smaller entropy than $S_{\text{eq}}$, corresponding to a nonequilibrium state. Notice that Equation (4) does not involve the governing equations of particle motion, e.g., Newton's laws. While the governing equations must be satisfied, since the system is chaotic, they offer no useful information of the probability of microstates.

In the models in [1-3], with the nonchaotic components, because the local particle behavior is dominated by kinetic theory, the governing equations play an explicit role in determining entropy. In other words, as the system is less random than the fully chaotic setup, it is possible to have more knowledge about $n_i$, which can be generally expressed as the additional constraints

$$\bar{f}_j(n_i) = 0 \ (j = 1,2 \ldots N_f) \tag{7}$$

where $\bar{f}_j$ indicate a set of $N_f$ differentiable functions. For example, in the two-shelf ideal-gas model in Figure 1(a,b) [1], there is one additional constraint ($N_f = 1$): $\bar{\rho} = \delta_1$, i.e., $\bar{f}_1 = \bar{\rho} - \delta_1 =$

---

0. The nonchaotic particle movements in the low-height step require that the expectation value of the particle number density ratio between the two shelves ($\bar{\rho}$) equals $\delta_1 = 1 - \mathrm{erf}(\sqrt{\beta m g z_s})$. For another example, in Figure 1(c) [2], the additional constraints may be written as $\bar{f}_j = \tilde{\rho}_j - \bar{\tau}_{\mathrm{pt}} \prod_{\chi=1}^{N_{\mathrm{tot}}} t_{j\chi} = 0$ ($j = 1,2 \ldots \Omega$), where $N_{\mathrm{f}} = \Omega$, $\tilde{\rho}_j$ is the probability of the $j$-th microstate, $t_{j\chi}$ is the time duration between the previous and next particle-wall collisions of the $\chi$-th particle in the $j$-th microstate, and $\bar{\tau}_{\mathrm{pt}}$ is the normalization factor.

With Equation (7), the Lagrangian is re-defined as

$$\mathcal{L} = k_{\mathrm{B}} \ln \Omega - \alpha_{\mathrm{n}}(\textstyle\sum_i n_i - N_{\mathrm{tot}}) - \beta_{\mathrm{n}}(\textstyle\sum_i n_i \epsilon_i - U) - \textstyle\sum_{j=1}^{N_{\mathrm{f}}} \gamma_j \bar{f}_j \tag{8}$$

where $\gamma_j$ are the additional Lagrange multipliers. Substituting Equation (8) into (4) gives

$$\frac{n_i}{g_i} = \varphi_{\mathrm{n}} \bar{\gamma}_i e^{-\beta \epsilon_i} \tag{9}$$

where $\bar{\gamma}_i = \exp(-\sum_{j=1}^{N_{\mathrm{f}}} k_{\mathrm{B}}^{-1} \gamma_j \bar{f}'_{ji})$ reflects the nonchaoticity effect, $\bar{f}'_{ji} = \partial \bar{f}_j / \partial n_i$, and $\varphi_{\mathrm{n}}$ is the normalization factor. Owing to $\bar{\gamma}_i$, Equation (9) differs from Equation (5), suggesting that the steady state is intrinsically nonequilibrium. It is noteworthy that Equation (7) is not in line with Boltzmann's assumption of equal *a priori* equilibrium probabilities. In Equation (9), at a given energy level $\epsilon_i$, $n_i$ varies with $\bar{\gamma}_i$.

Because Equation (9) is derived from Equation (4), entropy is maximized, i.e., the non-Boltzmann $n_i$ is compatible with the principle of maximum entropy. Combination of Equation (9) and Equation (2) gives the nonequilibrium entropy

$$S_{\mathrm{ne}} = k_{\mathrm{B}} N_{\mathrm{tot}} \ln \varphi_{\mathrm{n}}^{-1} + k_{\mathrm{B}} \beta U - k_{\mathrm{B}} \textstyle\sum_i (n_i \ln \bar{\gamma}_i) \tag{10}$$

With Equation (7), the more constrained maximization of entropy leads to the local maximum ($S_{\mathrm{ne}}$) in the phase space, smaller than the global maximum ($S_{\mathrm{eq}}$) at thermodynamic equilibrium.

Equations (6) and (10) can be collectively described by the generalized second law of thermodynamics [3]: in an isolated system, entropy cannot evolve away from the maximum possible value at the steady state ($S_{\mathrm{Q}}$), i.e.,

$$S \rightarrow S_{\mathrm{Q}} \tag{11}$$

In a fully chaotic system, $S_{\mathrm{Q}} = S_{\mathrm{eq}}$, and Equation (11) is equivalent to the entropy statement of the second law of thermodynamics. With the locally nonchaotic component, $S_{\mathrm{Q}} = S_{\mathrm{ne}}$, and entropy may spontaneously decrease from $S_{\mathrm{eq}}$ to $S_{\mathrm{ne}}$.

1.4 Questions to quantum statistical mechanics

Compared to classical mechanics [6], thermodynamic limit in quantum mechanics is relatively rarely reported. The investigation on the classical mechanical models in [1-3] (Sections 1.2 and 1.3) raises interesting questions for quantum models.

Quantum statistical mechanics has been extensively studied [e.g., 11]. In a quantum gas, in accordance with the innate rules of particle distribution [5],

$$\Omega = \begin{cases} \prod_i \frac{(n_i + g_i - 1)!}{n_i!(g_i - 1)!} & \text{(Bose gas)} \\ \prod_i \frac{g_i!}{n_i!(g_i - n_i)!} & \text{(Fermi gas)} \end{cases} \tag{12}$$

Substituting Equations (12) into (4) results in the Bose-Einstein distribution and the Fermi-Dirac distribution [5]:

$$\frac{n_i}{g_i} = \begin{cases} \frac{1}{Be^{\beta \epsilon_i} - 1} & \text{(Bose gas)} \\ \frac{1}{Be^{\beta \epsilon_i} + 1} & \text{(Fermi gas)} \end{cases} \tag{13}$$

where $B$ is determined by Equation (3). In the Fermi-Dirac distribution, $B$ is often written as $e^{-\beta \mu_F}$, with $\mu_F$ being the chemical potential. In the Bose-Einstein distribution, $B \geq 1$; when the particle number is not conserved (e.g., in blackbody radiation), $B = 1$.

Like Equation (9), Equation (13) is not the Maxwell-Boltzmann distribution (Equation 5), so that the associated entropy is unequal to $S_{eq}$ (Equation 6). However, unlike Equation (9), Equation (13) must not cause any conflict with the second law of thermodynamics; otherwise, quantum mechanics would inherently contradict statistical physics.

In Section 2, we examine the above concern. In Section 3, through a comparison of a set of classical mechanical and quantum mechanical single-particle models, the nature of quantum non-thermodynamic phenomena is further explored. Section 4 shows that the principle of maximum entropy is complied with.

A difficulty in extending research from classical mechanics [1-3] to quantum mechanics is that the fundamentals of quantum measurement (the collapse of the wave function) have not been adequately understood. The energy and information properties of quantum measurement are often unclear, somewhat similar to the situation of Maxwell's demon [8]. To circumvent this hurdle, our analyses will be based on the generalized Maxwell's relations (Sections 2 and 3) and the Gibbs entropy (Section 4). For the generalized Maxwell's relations, regardless of how the system state evolves on the microscopic scale, we observe the overall energy exchanges at the macroscopic/ensemble level; for the Gibbs entropy, the specifics of the wave function are treated as unknown.

## 2. Fermi gas and Bose gas

In this section, we reiterate that the Fermi-Dirac distribution and the Bose-Einstein distribution are inherently consistent with the heat-engine statement of the second law of thermodynamics, demonstrating the general robustness of the framework of quantum statistical mechanics. The mechanism underlying the difference between Equations (13) and (5) is unrelated to that of Equations (9) and (5). The gas particles are in thermal equilibrium, which will complement the discussion in Section 3.

### 2.1 Heat-engine statement of the second law of thermodynamics

The heat-engine statement of the second law of thermodynamics can be formulated as the balance of cross-influence of thermodynamic forces [1] (see Appendix 1 for the details):

$$\Delta_{12} = \Delta_{21} \tag{14}$$

where $\Delta_{12} \triangleq \partial F_1 / \partial x_2$, $\Delta_{21} \triangleq \partial F_2 / \partial x_1$, $F_1$ and $F_2$ are two thermally correlated thermodynamic forces, and $x_1$ and $x_2$ are their conjugate variables, respectively. Equation (14) may be viewed as the generalized Maxwell's relations [12]. Its various forms have been widely studied in many areas, such as the Nernst equation [13].

Figure 2(a) depicts a three-dimensional (3D) quantum gas. The system is closed and immersed in a thermal reservoir. It consists of the main zone and the high-potential zone, with the potential difference between them being denoted by $u$. The volume of the transition zone is negligible. A large number of gas particles randomly move. The total particle number ($N_{\text{tot}}$) and the high-potential zone volume ($V_0$) are constant. The main-zone volume ($V_{\text{m}}$) and the potential difference ($u$) can be adjusted independently. To increase $u$ (i.e., to raise the potential energy of the particles in the grey sphere) by a small amount $du$, the system consumes work $F_V dz$, where $F_V$ is the thermodynamic force and $z$ is the conjugate variable. For example, if $u$ is caused by gravity, $u = mgz$ and $F_V = mgN_{\text{H}}$, with $N_{\text{H}}$ being the particle number in the high-potential zone, and $z$ the height difference. For another example, if $u$ is caused by a voltage $\Delta V$, $z = \Delta V$, $u = e_0 \Delta V$, and $F_V = e_0 N_H$ is the total charge in the high-potential zone, where $e_0$ is the particle charge. Generally, we can define $F_0 = u/z$ to represent the contribution of one particle, since $F_V = F_0 N_H$. The high-potential zone is an embodiment of the internal thermodynamic processes. It is used as an auxiliary element in the thermodynamic analysis. Both $u$ and $V_0$ may be arbitrarily small, and neither of them will affect the equations of degeneracy pressure to be derived in the next section.

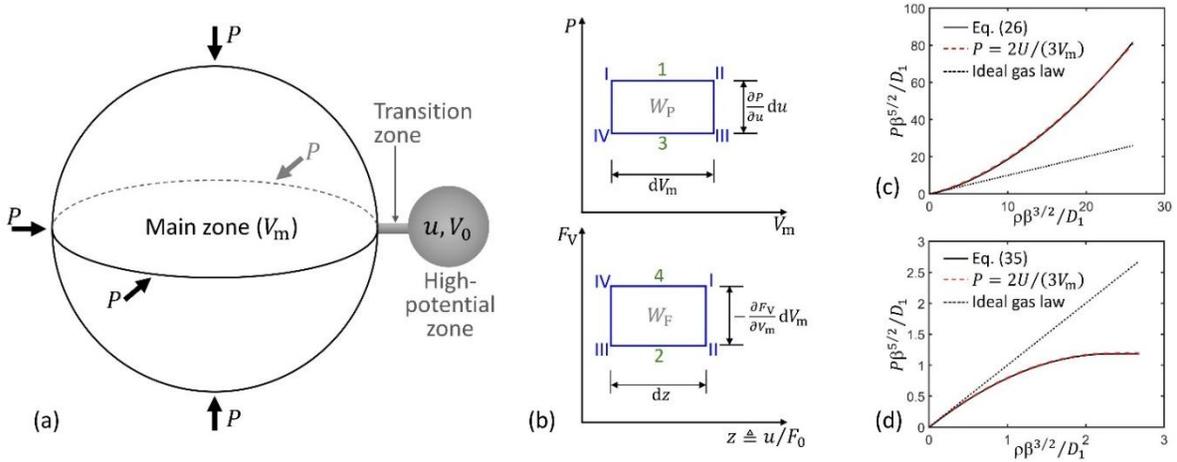

**Figure 2. (a)** Schematic of a three-dimensional quantum gas. The system is closed and immersed in a thermal reservoir. It consists of the main zone and the high-potential zone (the grey sphere); the volume of the transition zone between them is negligible. The high-potential zone is used as an auxiliary element in the thermodynamic analysis; $u$ and $V_0$ can be arbitrarily small and do not affect the derived equations of degeneracy pressure. **(b)** The four-step isothermal operation cycle. The degeneracy pressure of the main zone ($P$) and the support force of the high-potential zone ($F_V$) are operated alternately. Indexes I-IV indicate the system states; numbers 1-4 indicate the operation steps; $dV_m$ and $dz$ are arbitrarily small. **(c)** For Fermi gas and **(d)** Bose gas, the normalized degeneracy pressure ($P\beta^{5/2}/D_1$) is calculated as a function of the normalized particle number density ($\rho\beta^{3/2}/D_1$). The curves of Equations (26,35) overlap with the conventional result $P = 2U/(3V_m)$.

The conjugate variable of $P$ is $-V_m$. By setting $F_1 = P$ and $F_2 = F_V$ ($x_1 = -V_m$ and $x_2 = z$), Equation (14) becomes

$$\Delta_P = -\frac{1}{F_0}\Delta_F \tag{15}$$

where $\Delta_P \triangleq \partial P/\partial u$ and $\Delta_F \triangleq \partial F_V/\partial V_m$. For an equilibrium system, Equation (15) can be directly obtained by using the Helmholtz free energy $\mathcal{F}$. Because $P = -\partial \mathcal{F}/\partial V_m$ and $F_V = \partial \mathcal{F}/\partial z$, $F_0 \Delta_P = -\Delta_F = -\partial^2 \mathcal{F}/(\partial V_m \partial z)$. It could be understood through the isothermal cycle in Figure 2(b). Hereafter, the term "isothermal" is used to describe a reversible process wherein at every step, the expectation value of the particle kinetic energy ($\bar{K}$) remains the same as that of the environment (a thermal reservoir); if an operation causes $\overline{K}$ to vary, it will change back by exchanging heat with the environment before the next step is performed. In Figure 2(b), in step 1 (State I to II), $V_m$ is increased by a small volume $dV_m$. The system does work $P \cdot dV_m$ to the environment. As more particles are distributed in the main zone and less particles are in the high-potential zone, $F_V$ is reduced by $\Delta_F dV_m$. In Step 2 (State II to III), $u$ is lowered by a small amount $du$. The system does work $(F_V - \Delta_F dV_m)dz$ to the environment. As more particles can overcome the energy barrier of $u$ and less particles are in the main zone, $P$ is reduced by $\Delta_P$. In Step 3 (State III to IV), $V_m$ is compressed by $dV_m$. The system consumes work $(P - \Delta_P du)dV_m$, and $F_V$ increases by $\Delta_F dV_m$. Finally in Step 4 (State IV to I), $u$ is raised by $du$. The system consumes work $F_V dz$, and returns to the initial state. Overall, $P$ produces work $W_P = \Delta_P du dV_m$, and $F_V$ consumes work $W_V = -\Delta_F dV_m dz$. The heat-engine statement of the second law of thermodynamics requires that no useful work can be produced in a cycle by absorbing heat from the environment, i.e., $W_P = W_V$, which leads to $F_0 \Delta_P = -\Delta_F$ (Equation 15).

Equation (15) is not directly useful for the study of the equation of state. Denote the particle number in the main zone by $N$. In equilibrium, if the particle number density in the main zone ($\rho = N/V_m$) and temperature ($T$) are given, $P$ is determined. That is, $P = P(\rho, T)$, where $u$ implicitly comes in by influencing $\rho$. Consequently,

$$\frac{\partial P}{\partial u} = \frac{\partial P}{\partial \rho}\frac{\partial \rho}{\partial u} \tag{16}$$

With $F_V = F_0 N_H$ and $N_H = N_{tot} - N$, substitution of Equation (15) into Equation (16) gives

$$\frac{\partial P}{\partial \rho} = \frac{\frac{\partial N}{\partial V_m}}{\frac{\partial \rho}{\partial u}} \tag{17}$$

Equation (17) will be the basis of the thermodynamic analysis in Section 2.2. For reference, here we apply it to examine classical ideal gas. According to the Maxwell-Boltzmann distribution (Equation 5), $N = \overline{N}_c V_m \int_0^\infty e^{-\beta\epsilon_i} d\epsilon_i = \overline{N}_c V_m / \beta$ and $N_H = \overline{N}_c V_0 \int_u^\infty e^{-\beta\epsilon_i} d\epsilon_i = \overline{N}_c V_0 e^{-\beta u} / \beta$, where $\overline{N}_c$ is the normalization factor. Since $N_{tot} = N + N_H$, $\overline{N}_c = \beta N_{tot} / (V_m + e^{-\beta u} V_0)$ and $N = N_{tot} / (1 + e^{-\beta u} V_0 / V_m)$. Consequently, $\partial N / \partial V_m = \rho V_0 e^{-\beta u} / (V_0 e^{-\beta u} + V_m)$ and $\partial \rho / \partial u = V_m^{-1}(\partial N / \partial u) = \beta \rho V_0 e^{-\beta u} / (V_0 e^{-\beta u} + V_m)$, so that Equation (17) becomes $\partial P / \partial \rho = 1/\beta = k_B T$. Because $P = 0$ when $\rho = 0$, the gas pressure in the main zone is

$$P = \int_0^{N/V_m} \frac{\partial P}{\partial \rho} d\rho = \frac{N}{V_m} k_B T, \tag{18}$$

which is the ideal gas law, as it should be.

## 2.2 Degeneracy pressure of quantum gas

In this section, we derive the equations of degeneracy pressure of quantum gas by using Equation (17). The method is similar to that used for Equation (18).

### 2.2.1 Fermi gas

For a Fermi gas, the Fermi-Dirac distribution function is $f_{\text{FD}} = 1/(Be^{\beta\epsilon_i} + 1)$ (Equation 13). In the main zone, the density of states is $g_i = D_2 V_{\text{m}}\sqrt{\epsilon_i}$ [10], where $D_2 = s_0 D_1$, $s_0 = 2s + 1$, $s$ is spin, $D_1 = (2m)^{3/2}/(4\pi^2\hbar^3)$, $\hbar = h/(2\pi)$, and $h$ is the Planck constant. In the high-potential zone, $g_i = D_2 V_0\sqrt{\epsilon_i - u}$. The particle numbers in the main zone and the high-potential zone are, respectively,

$$N = \int_0^\infty g_i f_{\text{FD}} \mathrm{d}\epsilon_i = \int_0^\infty \frac{D_2 V_{\text{m}}\sqrt{\epsilon_i}}{Be^{\beta\epsilon_i}+1}\mathrm{d}\epsilon_i = D_{\text{B}} V_{\text{m}} \cdot Y(B) \tag{19}$$

$$N_{\text{H}} = \int_u^\infty g_i f_{\text{FD}} \mathrm{d}\epsilon_i = \int_u^\infty \frac{D_2 V_0\sqrt{\epsilon_i-u}}{Be^{\beta\epsilon_i}+1}\mathrm{d}\epsilon_i = D_{\text{B}} V_0 \cdot Y(Be^{\beta u}) \tag{20}$$

where $D_{\text{B}} = D_2/\beta^{3/2}$, and $Y(B) \triangleq \int_0^\infty [\sqrt{y}(1 + Be^y)^{-1}]\,\mathrm{d}y$. The total particle number is

$$N_{\text{tot}} = N + N_{\text{H}} = D_{\text{B}}\big[V_{\text{m}} \cdot Y(B) + V_0 \cdot Y(Be^{\beta u})\big] \tag{21}$$

On both sides of Equation (21), taking the derivative with respect to $V_{\text{m}}$ gives

$$\frac{\partial B}{\partial V_{\text{m}}} = \frac{Y(B)}{V_{\text{m}}\bar{Y}(B) + V_0 e^{\beta u} \cdot \bar{Y}(Be^{\beta u})}, \tag{22}$$

and taking the derivative with respect to $u$ gives

$$\frac{\partial B}{\partial u} = -\frac{V_0 B\beta e^{\beta u} \cdot \bar{Y}(Be^{\beta u})}{V_{\text{m}}\bar{Y}(B) + V_0 e^{\beta u} \cdot \bar{Y}(Be^{\beta u})}, \tag{23}$$

where $\bar{Y}(B) \triangleq \int_0^\infty [\sqrt{y}e^y(1 + Be^y)^{-2}]\mathrm{d}y$. According to Equation (19),

$$\frac{\partial \rho}{\partial u} = \frac{1}{V_{\text{m}}}\frac{\partial N}{\partial u} = -D_{\text{B}}\bar{Y}(B)\frac{\partial B}{\partial u} \tag{24}$$

According to Equation (20),

$$\frac{\partial N}{\partial V_{\text{m}}} = -\frac{\partial N_{\text{H}}}{\partial V_{\text{m}}} = D_{\text{B}} V_0 e^{\beta u}\bar{Y}(Be^{\beta u})\frac{\partial B}{\partial V_{\text{m}}} \tag{25}$$

Substitution of Equations (19,22-25) into Equation (17) leads to $\partial P/\partial \rho = \rho\sqrt{\beta}/[D_2 B \cdot \bar{Y}(B)]$. Since $P = 0$ when $\rho = 0$,

$$P = \int_0^{N/V_m} \frac{\partial P}{\partial \rho} d\rho = \frac{\sqrt{\beta}}{D_2} G_F(\rho, D_B) \tag{26}$$

where $G_F(\rho, D_B) = \int_0^{N/V_m} \rho[B \cdot \bar{Y}(B)]^{-1} d\rho$, and $B$ is given by $Y(B) = \rho/D_B$ (Equation 19). As shown in Figure 2(c), $P = \sqrt{\beta} G_F/D_2$ is equivalent to the conventional equation $P = 2U/(3V_m)$, where $U = \int_0^\infty f_{FD} g_i \epsilon_i d\epsilon_i$. Equation (26) may be approximated as (see Appendix 2)

$$P \approx (3\pi^2)^{2/3} \frac{\hbar^2}{5m} \rho^{5/3} + \mu_2 \left(\frac{4\pi^2}{s_0}\right)^{1/3} \frac{\hbar}{\sqrt{2m}} \rho^{4/3} \sqrt{k_B T} + \rho k_B T \tag{27}$$

where $\mu_2 \approx -0.35$.

### 2.2.2 Bose gas

For a Bose gas, the Bose-Einstein distribution function is $f_{BE} = 1/(Be^{\beta\epsilon_i} - 1)$ (Equation 13). For the sake of simplicity, we assume that $T$ is sufficiently high and no Bose-Einstein condensate is formed. In the main zone, $g_i = D_1 V_m \sqrt{\epsilon_i}$ [10]; in the high-potential zone, $g_i = D_1 V_0 \sqrt{\epsilon_i - u}$. The particle numbers in the main zone and the high-potential zone are

$$N = \int_0^\infty g_i f_{BE} d\epsilon_i = \int_0^\infty \frac{D_1 V_m \sqrt{\epsilon_i}}{e^{\beta(\epsilon_i - \mu)} - 1} d\epsilon_i = A_B V_m \cdot R(B) \tag{28}$$

$$N_H = \int_u^\infty g_i f_{BE} d\epsilon_i = \int_u^\infty \frac{D_1 V_0 \sqrt{\epsilon_i - u}}{e^{\beta(\epsilon_i - \mu)} - 1} d\epsilon_i = A_B V_0 \cdot R(Be^{\beta u}) \tag{29}$$

where $A_B = D_1/\beta^{3/2}$, and $R(B) \triangleq \int_0^\infty [\sqrt{y}(Be^y - 1)^{-1}] dy$. The total particle number is

$$N_{tot} = N + N_H = A_B[V_m \cdot R(B) + V_0 \cdot R(Be^{\beta u})] \tag{30}$$

On both sides of Equation (30), taking the derivative with respect to $V_m$ gives

$$\frac{\partial B}{\partial V_m} = \frac{R(B)}{V_m \bar{R}(B) + V_0 e^{\beta u} \bar{R}(Be^{\beta u})}, \tag{31}$$

and taking the derivative with respect to $u$ gives

$$\frac{\partial B}{\partial u} = -\frac{V_0 B \beta e^{\beta u} \bar{R}(Be^{\beta u})}{V_m \bar{R}(B) + V_0 e^{\beta u} \bar{R}(Be^{\beta u})}, \tag{32}$$

where $\bar{R}(B) \triangleq \int_0^\infty [\sqrt{y} e^y (Be^y - 1)^{-2}] dy$. According to Equations (28) and (29),

$$\frac{\partial \rho}{\partial u} = \frac{1}{V_m} \frac{\partial N}{\partial u} = -A_B \bar{R}(B) \frac{\partial B}{\partial u} \tag{33}$$

$$\frac{\partial N}{\partial V_{\mathrm{m}}} = -\frac{\partial N_{\mathrm{H}}}{\partial V_{\mathrm{m}}} = A_{\mathrm{B}} V_0 e^{\beta u} \bar{R}(Be^{\beta u}) \frac{\partial B}{\partial V_{\mathrm{m}}} \tag{34}$$

Substitution of Equations (31-34) into Equation (17) leads to $\partial P/\partial \rho = \rho \sqrt{\beta}/[D_1 B \cdot \bar{R}(B)]$. Since $P = 0$ when $\rho = 0$,

$$P = \int_0^{N/V_{\mathrm{m}}} \frac{\partial P}{\partial \rho} \mathrm{d}\rho = \frac{\sqrt{\beta}}{D_1} G_{\mathrm{B}}(\rho, A_{\mathrm{B}}) \tag{35}$$

where $G_{\mathrm{B}}(\rho, A_{\mathrm{B}}) = \int_0^{N/V_{\mathrm{m}}} \rho[B \cdot \bar{R}(B)]^{-1} \mathrm{d}\rho$, and $B$ is given by $R(B) = \rho/A_{\mathrm{B}}$ (Equation 28). As shown in Figure 2(d), Equation (35) is equivalent to the conventional equation $P = 2U/(3V_{\mathrm{m}})$, where the internal energy is $U = \int_0^\infty f_{\mathrm{BE}} g_i \epsilon_i \mathrm{d}\epsilon_i$.

### 2.2.3 Two-dimensional quantum gas

In Appendix 3, using the same approach as for the 3D cases, we derive the equations of degeneracy pressure of two-dimensional (2D) quantum gas:

$$P = \begin{cases} C_{\mathrm{B}} k_{\mathrm{B}} T \cdot P_{\mathrm{n}}(x_{\mathrm{u}}) & \text{(2D Bose gas)} \\ L_{\mathrm{B}} k_{\mathrm{B}} T \cdot P_{\mathrm{m}}(x_{\mathrm{v}}) & \text{(2D Fermi gas)} \end{cases} \tag{36}$$

where $P_{\mathrm{n}}(x_{\mathrm{u}}) = \mathrm{Li}_2(e^{x_{\mathrm{u}}}) + x_{\mathrm{u}} \ln(1 - e^{x_{\mathrm{u}}}) - x_{\mathrm{u}}^2/2 - \pi^2/6$, $x_{\mathrm{u}} = \rho/C_{\mathrm{B}}$, $C_{\mathrm{B}} = m/(2\beta\pi\hbar^2)$, $P_{\mathrm{m}}(x_{\mathrm{v}}) = P_{\mathrm{n}}(x_{\mathrm{v}}) + x_{\mathrm{v}}^2/2 = \mathrm{Li}_2(e^{x_{\mathrm{v}}}) + x_{\mathrm{v}} \ln(1 - e^{x_{\mathrm{v}}}) - \pi^2/6$, $x_{\mathrm{v}} = \rho/L_{\mathrm{B}}$, and $L_{\mathrm{B}} = s_0 C_{\mathrm{B}}$. It can be verified that Equation (36) is equivalent to the conventional equation $P = U/A$, where $A$ indicates area.

### 2.3 Consistency with the second law of thermodynamics

The equations of degeneracy pressure of quantum gas (Equations 26, 35, and 36) are derived from Equations (14) and (15). As expected, they depend only on the parameters of the main zone ($m$, $T$, and $\rho$) and are in agreement with $P = 2U/(3V_{\mathrm{m}})$ (for 3D gas) or $P = U/A$ (for 2D gas). This result confirms that in thermal equilibrium, a quantum gas intrinsically obeys the heat-engine statement of the second law of thermodynamics. While the distribution of the quantum particles is not the same as the Maxwell-Boltzmann distribution, the equations of degeneracy pressure also differ from the ideal gas law. The two effects counterbalance each other, so that both classical gas and quantum gas fit with the theory of statistical mechanics.

As for the principle of maximum entropy (Equation 4), for quantum gas, it is the Fermi-Dirac distribution (for a Fermi gas) or the Bose-Einstein distribution (for a Bose gas) that drives $S$ to reach the maximum possible value, not the Maxwell-Boltzmann distribution. Equation (13) is in line with Boltzmann's assumption of equal *a priori* equilibrium probabilities. It does not contain any nonequilibrium element independent of $\epsilon_i$. Contrariwise, in classical mechanics, the nonchaoticity-induced additional constraints (Equation 7) do not affect $S_{\mathrm{eq}}$. Instead, Equation (7) reduces entropy to the local maximum in the phase space ($S_{\mathrm{ne}}$). The mismatch between $S_{\mathrm{eq}}$ and $S_{\mathrm{ne}}$ renders the steady state intrinsically nonequilibrium.

## 3. Models of simple-step scattering

The study in [1-3] analyzed a few classical mechanical models with locally nonchaotic barriers (e.g., Figure 1) and demonstrated that their steady states are intrinsically nonequilibrium, incompatible with the second law of thermodynamics. In this section, we investigate the quantum counterpart of the energy-barrier model, focusing on the simplest possible one-dimensional (1D) single-particle setups that have well-established analytical solutions.

The analysis will be based on the heat-engine statement of the second law of thermodynamics, formulated as the generalized Maxwell's relations (Equation 14). It examines the time-averaged energy exchanges at the ensemble level, without necessarily relying on the detailed form of the wave function.

### 3.1 Classical mechanical models

We first compare a few classical mechanical cases. Figure 3(a) depicts a single particle moving along the $x$-axis in a 1D box. The box width is denoted by $x_0$. The potential at the outer boundaries ($x = 0$ and $x = x_0$) is infinitely large. Inside the box, there is a potential barrier $V$. The barrier separates the box into the upper shelf and the lower shelf, with the widths being denoted by $a$ and $b$, respectively. Both $a$ and $b$ are much larger than $z = V/F_0$, and $a + b = x_0$. The lower-shelf width ($b$) and the potential barrier ($V$) can be adjusted independently, and $x_0$ is fixed.

There are two thermodynamic forces: the force required to change $b$ ($F_b$) and the force required to change $V$ ($F_v$). The point of action of $F_b$ is at the energy barrier ($x = b$), and $F_v$ is applied on the upper shelf. For $F_b$ and $F_v$, Equation (14) becomes

$$\Delta_{vb} = -F_0 \Delta_{bv} \tag{37}$$

where $\Delta_{vb} \triangleq \partial F_v / \partial b$ and $\Delta_{bv} \triangleq \partial F_b / \partial V$. When Equation (37) is satisfied, no useful work can be produced in the operation cycle in Figure 3(b), similarly to Figure 2(b). In Figure 3(b), from State i to ii, $F_b$ increases $b$ by a small length $db$, accompanied by a decrease in $F_v$ of $|\Delta_{vb}|db$. From State ii to iii, $F_v$ increases $z$ by a small amount $dz = dV/F_0$, accompanied by a variation in $F_b$ by $\Delta_{bv}dV$. From State iii to iv, $b$ is reduced by $db$, and $F_v$ changes back. From State iv to i, $V$ is decreased by $dV$, and the system returns to the initial state. In a complete cycle, $F_v$ produces work $W_v = (|\Delta_{vb}|db)dz$, and $F_b$ consumes work $W_b = (|\Delta_{bv}|dV)db$. The heat-engine statement of the second law of thermodynamics dictates that $W_v = W_b$, which leads to $\Delta_{vb} = -F_0\Delta_{bv}$ (Equation 37).

In the current study, $F_v$ and $F_b$ are analyzed as the expectation values. Figure 3(c) depicts an ensemble of an arbitrarily large number of single-particle boxes. The boxes are operated simultaneously and always have the same $b$ and the same $V$. The microstates of the boxes are random and independent of each other, and $F_v$ and $F_b$ are the ensemble-average forces.

### 3.1.1 Perfectly insulated setup

In this section, we assume that the box in Figure 3(a) is thermally insulated, and show that such a system can be described by thermodynamics. The outer boundaries ($x = 0$ and $x = x_0$) are specular walls. The ensemble-average $F_v$ in Figure 3(c) is

$$F_v = \begin{cases} F_0 \dfrac{t_u}{t_l + t_u} & (E > V) \\ 0 & (E \le V) \end{cases} \tag{38}$$

where $E$ is the particle energy, and $t_l = b/\sqrt{2E/m}$ and $t_u = a/\sqrt{2(E-V)/m}$ are the characteristic travel times on the lower and upper shelves, respectively. In accordance with the conservation of momentum, the ensemble-average $F_b$ is

$$F_b = \begin{cases} \dfrac{\sqrt{2mE} - \sqrt{2m(E-V)}}{t_b} \dfrac{t_b}{t_l + t_u} = \dfrac{\sqrt{2mE} - \sqrt{2m(E-V)}}{t_l + t_u} & (E > V) \\ \dfrac{2\sqrt{2mE}}{t_b} \dfrac{t_b}{2t_l} = \dfrac{\sqrt{2mE}}{t_l} & (E \le V) \end{cases} \tag{39}$$

where $t_b$ indicates the characteristic time of the particle-barrier interaction. When $E \leq V$, since $\Delta_{vb} = \Delta_{bv} = 0$, Equation (37) is trivially satisfied. When $E > V$, based on Equations (38,39),

$$\Delta_{vb} = \Phi_F \left[ \frac{\sqrt{E/V} - \sqrt{E/V - 1}}{\sqrt{(E/V - 1)(E/V)}} \right] - \Phi_G = -\frac{\Phi_F}{\sqrt{E/V}} \frac{x_0/b}{x_0/b - 1} \tag{40}$$

$$\Delta_{bv} = \frac{\Phi_F}{F_0} \left[ \Phi_G - \frac{\sqrt{E/V} - \sqrt{E/V - 1}}{E/V - 1} \right] \tag{41}$$

where $\Phi_F = (F_0/b)(\phi_f x_0/b - 1)[\phi_g + \phi_f(x_0/b - 1)]^{-2}$, $\Phi_G = \phi_g/(x_0/b - 1) + \phi_f$, $\phi_g = 1/\sqrt{E/V}$, and $\phi_f = 1/\sqrt{E/V - 1}$. Equations (40,41) suggest that $-F_0 \Delta_{bv} > \Delta_{vb}$ for all $E/V > 1$. Since $\Delta_{vb} \neq -F_0 \Delta_{bv}$, the operation cycle in Figure 3(b) can produce work $\Delta W = W_v - W_b$.

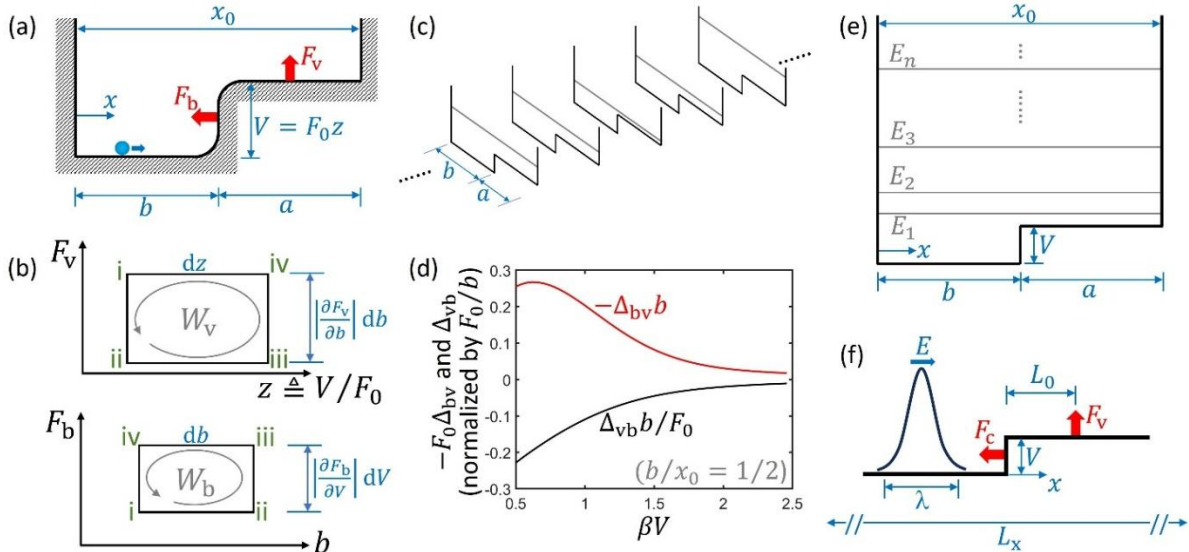

**Figure 3** (a) One-dimensional classical mechanical model of a particle in a box. There is a potential barrier ($V$) at $x = b$. (b) The four-step operation cycle. Indexes i-iv indicate the system states; $dz$ and $db$ are arbitrarily small. (c) An ensemble of a large number of single-particle boxes. All the boxes are operated simultaneously, and always have the same $b$ and the same $V$. We analyze the ensemble-average forces ($F_v$ and $F_b$). (d) Comparison between $\Delta_{vb}$ and $-F_0 \Delta_{bv}$ as functions of $\beta V$; $b/x_0$ is set to 1/2. (e) Quantum mechanical models of an infinite potential well and (f) a free particle. The particle interacts with the potential barrier ($V$).

The work production ($\Delta W$) is allowed by the second law of thermodynamics. It comes from the particle energy ($E$) and after a cycle, the initial state is not restored. Moreover, in each single-particle box in Figure 3(c), while the particle position is random, $E$ is the same. To prepare the ensemble, we need to measure $E$, and $\Delta W$ may not compensate the energetic penalty of information processing, similarly to the Szilárd engine [8].

### 3.1.2 Perfectly isothermal setup

In this section, the setup in Figure 3(a) is fully immersed in a thermal reservoir; we show that such a system follows the second law of thermodynamics. The particle can exchange heat with the environment. Assume that the system is perfectly isothermal. That is, at any position $x$ in Figure 3(c), the particle speed ($v$) always follows the 1D Maxwell-Boltzmann distribution $p(v) = \sqrt{2\beta m/\pi}\, e^{-\beta m v^2/2}$.

Across the energy barrier, the average particle flux per unit time from the lower shelf to the upper shelf is

$$\bar{J}_1 = \rho_{\mathrm{L}} \int_{\sqrt{2V/m}}^{\infty} [v \cdot p(v)] \mathrm{d}v = \rho_{\mathrm{L}} e^{-\beta V} \sqrt{\frac{2}{\pi \beta m}}, \qquad (42)$$

and in the inverse direction

$$\bar{J}_2 = \rho_{\mathrm{U}} \int_{0}^{\infty} [v \cdot p(v)] \mathrm{d}v = \rho_{\mathrm{U}} \sqrt{\frac{2}{\pi \beta m}}, \qquad (43)$$

where $\rho_{\mathrm{L}}$ and $\rho_{\mathrm{U}}$ are the probability densities for the particle to be on the lower and upper shelves, respectively. At the steady state, $\bar{J}_1 = \bar{J}_2$, so that

$$\frac{\rho_{\mathrm{U}}}{\rho_{\mathrm{L}}} = e^{-\beta V} \qquad (44)$$

Since $\rho_{\mathrm{U}}/\rho_{\mathrm{L}}$ is the Boltzmann factor, thermodynamic equilibrium is reached. Under this condition, Equation (37) is balanced [1], in agreement with the heat-engine statement of the second law of thermodynamics.

### 3.1.3 Mixed boundary condition: a non-thermodynamic system

In this section, we show that if the particle exchanges heat with the environment only at the outer boundaries, the system cannot reach thermal equilibrium and its behavior may conflict with the second law of thermodynamics. The borders at $x = 0$ and $x = x_0$ are thermal walls, and everywhere else ($0 < x < x_0$) is thermally insulated. When the particle collides with the thermal wall at $x = 0$ or $x = x_0$, the reflected particle speed is not correlated with the incident speed; instead, it randomly follows the 1D Maxwell-Boltzmann distribution $p(v)$.

In Figure 3(c), use $J_{\mathrm{U}}$ and $J_{\mathrm{L}}$ to denote the numbers of the particles that arrive at the energy barrier per unit time on the upper and lower shelves, respectively. At the steady state, their ratio is

$\bar{J}_x \triangleq J_U/J_L = \int_{\sqrt{2V/m}}^{\infty} p(v)\mathrm{d}v = 1 - \mathrm{erf}(\beta V)$ . After the particle has either crossed or been reflected by the energy barrier, the average outgoing speed ($\bar{v}_r$) away from the barrier is different from $\bar{v}_0 = \int_0^{\infty} v \cdot p(v)\mathrm{d}v = \sqrt{2/(\pi\beta m)}$. On the upper shelf, $\bar{v}_r$ is

$$\bar{v}_1 = \int_{\sqrt{2V/m}}^{\infty} \sqrt{v^2 - \frac{2V}{m}} \, p(v)\mathrm{d}v; \tag{45}$$

on the lower shelf, $\bar{v}_r$ is

$$\bar{v}_2 = (1 - \bar{J}_x) \int_0^{\sqrt{2V/m}} v \cdot p(v)\mathrm{d}v + \bar{J}_x \int_0^{\infty} \sqrt{v^2 + \frac{2V}{m}} \, p(v)\mathrm{d}v. \tag{46}$$

Overall, between the upper and lower shelves, the ratio of the probability of finding the particle is

$$\bar{\rho}_x = \frac{\rho_U}{\rho_L} = \bar{J}_x \frac{a(1/\bar{v}_0 + 1/\bar{v}_1)}{b(1/\bar{v}_0 + 1/\bar{v}_2)} \tag{47}$$

Thus, the ensemble-average $F_v$ is

$$F_v = F_0 \frac{\bar{\rho}_x}{1 + \bar{\rho}_x} = \frac{F_0(x_0 - b)(\bar{v}_0 + \bar{v}_1)}{(x_0 - b)(\bar{v}_0 + \bar{v}_1) + \frac{b(\bar{v}_0 + \bar{v}_2)\bar{v}_1}{[1 - \mathrm{erf}(\beta V)]\bar{v}_2}} \tag{48}$$

The ensemble-average $F_b$ can be assessed as

$$F_b = \frac{m(\bar{v}_0 - \bar{v}_2)}{\Delta t_b} \frac{\Delta t_b}{b(1/\bar{v}_0 + 1/\bar{v}_2)} - \bar{\rho}_x \frac{m(\bar{v}_0 - \bar{v}_1)}{\Delta t_b} \frac{\Delta t_b}{a(1/\bar{v}_0 + 1/\bar{v}_1)} \tag{49}$$

$$= \frac{m\bar{v}_0 \bar{v}_2}{b(\bar{v}_0 + \bar{v}_2)} \{\bar{v}_0 \, \mathrm{erf}(\beta V) + [1 - \mathrm{erf}(\beta V)]\bar{v}_1 - \bar{v}_2\}$$

The first term at the right-hand side of Equation (49) is for the lower shelf; effectively, at the steady state, the particles on the lower shelves in Figure 3(c) collide with the energy barrier at $\bar{v}_0$ and are "reflected" at $\bar{v}_2$. The second term is for the upper shelf; effectively, at the steady state, the particles on the upper shelves collide with the energy barrier at $\bar{v}_0$ and are "reflected" at $\bar{v}_1$.

As detailed in Appendix 4 and shown in Figure 3(d), Equations (48,49) suggest that Equation (37) is unbalanced ($\Delta_{vb} \neq -F_0 \Delta_{bv}$), i.e., the heat-engine statement of the second law of thermodynamics is not adhered to. This counterintuitive phenomenon is compatible with the discussion of Equations (40, 41) that $-F_0 \Delta_{bv} > \Delta_{vb}$ holds true for all $v > \sqrt{2V/m}$ (i.e., $E > V$), and the range of $v \leq \sqrt{2V/m}$ (i.e., $E \leq V$) is trivial; therefore, in the entire spectrum of $v$, $-F_0 \Delta_{bv}$ cannot be equal to $\Delta_{vb}$.

Figure 3(c) contains many particles. The operation of the thermodynamic forces ($F_b$ and $F_v$) is pre-programmed, without any knowledge of the detailed microstate of each box. The inconsistency with Equation (37) cannot be explained by the physical nature of information.

In essence, the mixed boundary condition results in an intrinsically nonequilibrium steady state ($\bar{\rho}_x \neq e^{-\beta V}$). The setup is a variant of the locally nonchaotic energy-barrier model in Figure 1(a,b). As the latter is not compatible with the second law of thermodynamics [1,2], it is unsurprising that the former contradicts Equation (37), allowing for production of useful work by absorbing heat from the environment with no other effect.

Indeed, as demonstrated by Maxwell's double-column engine [14], the very fact that the system cannot relax to thermal equilibrium ($\bar{v}_1 \neq \bar{v}_2$) already renders the second law of thermodynamics inapplicable. In Figure 3(c), as $\bar{v}_1 > \bar{v}_2$, the particles spontaneously and continuously transport heat from the thermal walls at the lower shelf to the thermal walls at the upper shelf, contradicting the refrigeration statement of the second law of thermodynamics. This process is equivalent to Figure 1(c) [2], with two thermally insulated constant-potential shelves attached to the high-potential and low-potential sides of the energy barrier, respectively. The constant-potential shelves do not alter the intrinsically nonequilibrium nature of the thermal state.

## 3.2 Quantum particle in a box: bound state

Figure 3(e) depicts the quantum counterpart of Figure 3(a), a potential barrier ($V$) at $x = b$ in an infinite square well. Equation (37) remains relevant, since it is derived from the heat-engine statement of the second law of thermodynamics, regardless of whether the system is classical mechanical or quantum mechanical. Consider the case wherein the potential barrier is much less than the expectation value of the particle energy, so that the particle state is dominated by $E > V$. As will be shown in Section 3.3, the condition of interest to us is when $E$ is large and the gaps among the energy eigenvalues are small.

The solution of the time-independent Schrödinger equation of this type of problem has been well established [15]. The eigen values of $E$ are (see Appendix 5)

$$E_n = V + \frac{\hbar^2}{2mx_0^2}(n\pi - \operatorname{atan}\alpha_D)^2 \quad (n = 1,2,3 \ldots) \tag{50}$$

where $\alpha_D = (\alpha_k k_2 \cos k_2 b - \sin k_2 b)/(\alpha_k k_2 \sin k_2 b + \cos k_2 b)$, $\alpha_k = \tan(k_1 b)/k_1$, $k_1 = \sqrt{2mE}/\hbar$, and $k_2 = \sqrt{2m(E-V)}/\hbar$. When $V \to 0$, $\alpha_D \to 0$ and $E_n$ is reduced to $\bar{E}_n = (n\pi\hbar)^2/(2mx_0^2)$. Numerical experiments on Equation (50) suggest that when $E_n \gg V$ (e.g., $n \geq 6$),

$$E_n \approx \bar{E}_n + \frac{a}{x_0}V. \tag{51}$$

If the particle is in an eigenstate $|\psi_n\rangle$, $E$ is $E_n$. According to Equation (50), when $V$ is changed by an arbitrarily small amount $dV$ ($F_v dz$ is much less than the difference between $E_n$ and $E_{n\pm1}$, $\delta E_n$), $F_v/F_0 = \partial E_n/\partial V$; when $b$ is changed by an arbitrarily small amount $db$ ($F_b db$ is much less than $\delta E_n$), $F_b = -\partial E_n/\partial b$. Because

$$-F_0 \Delta_{bv} = -F_0 \frac{\partial F_b}{\partial V} = F_0 \frac{\partial^2 E_n}{\partial b \partial V} \tag{52}$$

$$\Delta_{vb} = \frac{\partial F_v}{\partial b} = F_0 \frac{\partial^2 E_n}{\partial V \partial b}, \tag{53}$$

Equation (37) is inherently satisfied, i.e., $-F_0 \Delta_{bv} = \Delta_{vb}$. In contrast to Section 3.1.1, the energetic penalty of information processing does not need to be accounted for.

If the particle is in a pure state $|\psi\rangle = \sum_n c_n |\psi_n\rangle$, the expectation value of energy is $\langle H \rangle = \sum_n |c_n|^2 E_n$, with $c_n$ being the coefficient of $|\psi_n\rangle$. For the ensemble in Figure 3(c), since

$$F_v = F_0 \frac{\partial \langle H \rangle}{\partial V} = F_0 \sum_n |c_n|^2 \frac{\partial E_n}{\partial V} \tag{54}$$

$$F_b = -\frac{\partial \langle H \rangle}{\partial b} = -\sum_n |c_n|^2 \frac{\partial E_n}{\partial b}, \tag{55}$$

we also have Equation (37):

$$-F_0 \Delta_{bv} = \Delta_{vb} = F_0 \sum_n \left( |c_n|^2 \frac{\partial^2 E_n}{\partial b \partial V} \right). \tag{56}$$

If the particle is in a mixed state, in a box in Figure 3(c), the particle is in a possible state $|\psi_k\rangle$, with a certain probability $p_k$. As shown by Equation (56), every box follows $\Delta_{vb} = -F_0 \Delta_{bv}$. Consequently, the entire ensemble satisfies Equation (37).

The above analyses suggest that in general, when energy is quantized, the system has a strong tendency to adhere to the second law of thermodynamics. Unlike the classical mechanical models in Section 3.1, the details of boundary condition, such as whether the insulation or isothermal condition is perfect, do not play an active role. There is no need to resort to the physical nature of information to avoid production of useful work from a single thermal reservoir. This characteristic is rooted in the nonlocal nature of the wave function. The produced or consumed work ($\Delta W$) is fully reflected by the change in $E_n$ or $\langle H \rangle$. That is, $E_n$ or $\langle H \rangle$ may be effectively used as the free energy, with Equation (37) being its innate "hallmark".

### 3.3 Quantum free particle: scattering state

The stark difference between Sections 3.1 and 3.2 begs a critical question: in the framework of Bohr's correspondence principle, what is the connection between the quantum mechanical state and the classical mechanical state? The former appears to inherently obey the second law of thermodynamics, whereas the latter may not [1-3].

The answer may lie in the scattering states. Figure 3(f) depicts one example. At $x = 0$, there is a potential barrier $V$. The effective system size ($L_x$) is much larger than the thermal wavelength $\lambda = h/\sqrt{2\pi m k_B T}$, so that the setup can be modeled as a free particle (see Appendix 6). Similarly to Section 3.2, $V$ is much less than the expectation value of the particle energy. The solution of the time-independent Schrödinger equation is given in Appendix 7 [16].

Use $F_c$ to denote the force required to move the potential barrier along the $x$-axis. In Figure 3(c), when the number of the units in the ensemble is much larger than $L_x/\lambda$, the ensemble-average $F_c$ can be statistically defined. For $F_v$ and $F_c$, Equation (14) becomes

$$\Delta_{vc} = -F_0 \Delta_{cv} \tag{57}$$

where $\Delta_{vc} \triangleq \partial F_v/\partial x_c$, $\Delta_{cv} \triangleq \partial F_c/\partial V$, and $x_c$ is the conjugate variable of $F_c$. The wave packet depends on the initial condition. Since $E$ is not quantized, computation of $F_v$ and $F_c$ cannot be based on Equations (52-56). Nevertheless, for a canonical ensemble, when the thermal wavelength ($\lambda$) is larger than the distance between the points of application of $F_v$ and $F_c$ ($L_0$), it is reasonable to assume that the system follows the second law of thermodynamics, because the wave function is effectively nonlocal across $L_0$. As the mixed boundary condition (Section 3.1.3) is irrelevant, the scenario is comparable to those of Section 3.1.2 and Section 3.2, which have no internal source of nonequilibrium phenomena.

When $\lambda \ll L_0$, however, the system is "semiclassical", in the sense that the detailed profile of wave packet is relatively insignificant. The particle could be subject to various boundary conditions at different $x$. If a portion of the boundaries is thermally insulated while the rest sections are thermal walls, the imperfect isothermal condition can spontaneously cause the particle energy distribution to be nonuniform, and the probability density may not be proportional to the Boltzmann factor. It may contradict Equation (57), for the same reason as in Section 3.1.3. If $\lambda$ is sufficiently small, the quantum model is reduced to the classical mechanical setup.

In other words, as $\lambda \ll L_0$, the wave function is not nonlocal. The speed of wave packet ($v_w$) can be defined. Without particle-particle interaction to homogenize the overall state, the system lacks a mechanism to reach global equilibrium. For example, by using the mixed boundary condition in Section 3.1.3, the outer borders of the high-potential and low-potential sections are thermal walls, and the interior is thermally insulated. The average particle energy in the high-potential section is unequal to that of the low-potential section, because of the potential barrier $V$. Hence, in Figure 3(c), the particles can transport heat between the thermal walls of the two sections in an unforced manner, as in the classical mechanical models in Figure 1(c) and Section 3.1.3. Furthermore, the characteristic frequency of particle-wall collisions ($f_w$) is dependent on $v_w$, which will be discussed in Section 4 below.

In such a setup, the nonchaotic nature of the individual particle behavior is responsible for the out-of-equilibrium steady state. The model fundamentally differs from Feynman's ratchet [9] or its variants, such as the single-electron refrigerator [17], the quantum "perpetuum mobile" [18,19], and the Meissner effect [20,21]. The contrast between the global wave function in Figure 3(e) and the localized wave packet in Figure 3(f) may be compared to the study in [22,23].

## 4. Consistency with the principle of maximum entropy

Section 1.3 demonstrates how, in classical mechanics, while the steady state of a locally nonchaotic system is intrinsically nonequilibrium, entropy is still maximized. In this section, the investigation is extended to quantum mechanics. As explained in Appendix 8, for a general analysis and to circumvent the complexity of quantum measurement, our discussion is based on the Gibbs entropy.

With the probability of the $k$-th possible state ($|\psi_k\rangle$) being denoted by $p_k$, the density matrix of a mixed state is

$$\hat{\rho} = \sum_k p_k |\psi_k\rangle\langle\psi_k| \tag{58}$$

In the single-particle model in Figure 3, $|\psi_k\rangle$ are an orthogonal basis. To evaluate the applicability of the second law of thermodynamics, the critical issue is not how to determine the probability distribution or the expectation value of the measurement result of any observable, but whether the steady state is in equilibrium or nonequilibrium, i.e., whether $p_k$ is proportional to the Boltzmann factor $\delta_0 = e^{-\beta \epsilon_k}$, with $\epsilon_k$ being the energy level of $|\psi_k\rangle$. To analyze $p_k$, the details of $|\psi_k\rangle$ are relatively nonessential.

When the system is in contact with a thermal reservoir, usually, it is assumed that thermal equilibrium can be reached and $p_k \propto \delta_0$. The Gibbs entropy is

$$S = -k_B \sum_k p_k \ln p_k \tag{59}$$

If $p_k$ is only subject to the constraints of

$$\sum_k p_k = 1 \text{ and } \sum_k p_k \epsilon_k = U, \tag{60}$$

the Lagrangian is

$$\mathcal{L} = -k_B \sum_k p_k \ln p_k - \alpha_h(\sum_k p_k - 1) - \beta_h(\sum_k p_k \epsilon_k - U), \tag{61}$$

where $\alpha_h$ and $\beta_h$ are the Lagrange multipliers. To maximize $S$, through

$$\frac{\partial \mathcal{L}}{\partial p_k} = 0, \tag{62}$$

we have

$$p_k \propto e^{-\beta \epsilon_k}, \tag{63}$$

where $\beta = \beta_h/k_B$. Unlike the density matrix and the von Neumann entropy [24], the Gibbs entropy (Equation 59) is independent of $|\psi_k\rangle$. That is, $|\psi_k\rangle$ is treated as unknown.

When $\lambda \ll L_0$, the wave packet in Figure 3(f) may encounter various local boundary conditions along the $x$-axis, without particle-particle interaction. Every time when the particle collides with a thermal wall, effectively, a quantum "measurement" takes place and the particle energy ($E$) is changed by the environment (a thermal reservoir). Between two "measurement" events, the wave function is governed by the Schrödinger equation. The frequency of particle-boundary collisions ($f_w$) is dependent on the speed of the wave packet ($v_w$), which in turn reflects $E$. As $f_w$ represents the rate at which $E$ and $v_w$ are updated, $p_k$ is influenced by the kinetic characteristics of $|\psi_k\rangle$, which may be generally described as

$$\bar{F}_j(p_k, \psi_k) = 0 \ (j = 1, 2 \dots n_f) \tag{64}$$

with $\bar{F}_j$ indicating a set of $n_f$ functions derived from kinetic theory. Equation (64) is the quantum counterpart of Equation (7) and is distinct from Equation (63). The Lagrangian is redefined as

$$\mathcal{L} = -k_B \sum_k p_k \ln p_k - \alpha_z(\sum_k p_k - 1) - \beta_z(\sum_k p_k \epsilon_k - U) - \sum_{j=1}^{n_f} \Gamma_j \bar{F}_j \tag{65}$$

where $\alpha_z$, $\beta_z$, and $\Gamma_j$ are the Lagrange multipliers. Compared to Equation (61), Equation (65) has the extra terms of $\sum_{j=1}^{n_f} \Gamma_j \bar{F}_j$, and maximization of $S$ (Equation 62) requires

$$k_B \ln p_k + \sum_{j=1}^{n_f} \Gamma_j \frac{\partial \bar{F}_j}{\partial p_k} = -k_B - \alpha_z - \beta_z \epsilon_k. \tag{66}$$

The solution of Equation (66) is different from that of Equation (61), i.e., $p_k$ is non-Boltzmannian. Since $v_w$ affects $p_k$, the system may never relax to thermal equilibrium, contrary to the quantum gas models in Section 2.

For instance, assume that the boundary setting in Figure 3(f) is the same as in Section 3.1.3. The outer borders at $x = -L_x/2$ and $x = L_x/2$ are thermal walls ($L_x \gg \lambda$), and everywhere else in between ($-L_x/2 < x < L_x/2$) is thermally insulated. In Figure 3(c), because the fast particles rapidly collide with the thermal walls and change particle energy, it is less likely to find them in the ensemble, compared to the equilibrium state; vice versa. This phenomenon is similar to the classical mechanical example in Section 1.1 ($\bar{p}_v \propto p_{MB}/v$) [2]. If $V = 0$, $f_w \propto v_w$. When $V > 0$, as $v_w$ is determined by $\epsilon_k$, $f_w$ is positively corelated with $\epsilon_k$, which can be denoted by $f_w = \bar{f}_v(\epsilon_k)$, with $\bar{f}_v$ being a positive-correlation function. Equation (64) becomes

$$\bar{F}_j(p_k, \psi_k) = \bar{f}_v(\epsilon_k) - f_w = 0 \ (j = 1, 2 \dots \Omega), \tag{67}$$

Accordingly, $p_k$ is inversely proportional to $\bar{f}_v(\epsilon_k)$. The overall effect may be expressed as

$$p_k = \bar{\tau}_k \frac{e^{-\beta\epsilon_k}}{\bar{f}_v(\epsilon_k)} \tag{68}$$

where $\bar{\tau}_k$ is the normalization factor ($\sum_k p_k = 1$). Equation (68) differs from Equation (63), i.e., the steady state is intrinsically nonequilibrium.

Since Equation (68) is derived from $\partial\mathcal{L}/\partial p_k = 0$, the associated Gibbs entropy $S_{ne} = -k_B \sum_k p_k \ln p_k = -k_B(\ln \bar{\tau}_k + \bar{P}_s)$ is maximized, where $\bar{P}_s = \sum_k [p_k \ln(e^{-\beta\epsilon_k}/\bar{f}_v)]$. With the additional constraints of $p_k \propto 1/\bar{f}_v$, the maximized entropy ($S_{ne}$) is not the global maximum in the phase space (the equilibrium entropy $S_{eq}$), but instead the local maximum in the region defined by Equation (67). In general, $S_{ne} < S_{eq}$. If initially the system is in equilibrium (i.e., $p_k \propto e^{-\beta\epsilon_k}$), as it evolves toward the intrinsically nonequilibrium steady state (i.e., $p_k \propto e^{-\beta\epsilon_k}/\bar{f}_v$), $S$ decreases from $S_{eq}$ to $S_{ne}$.

The out-of-equilibrium characteristic of Equation (68) is innate, not induced by any externally applied thermodynamic driving force. The thermal-wall boundary condition reflects the equilibrium state of the environment (a thermal reservoir). When the particle interacts with a thermal wall, the quantum "measurement" happens spontaneously, and no work is consumed.

If the environment is not a thermal reservoir but a nonchaotic medium, even when energy is quantized, there could be additional constraints other than Equation (60), and $p_k \propto \delta_0$ may not hold true. Notice that Figure 3(f) is unrelated to many-body localization (MBL). MBL studies a subregion of interacting particles with known initial and boundary conditions [25]. If the initial state is random, the expectation value of an observable would agree with quantum statistical mechanics [26]. On the contrary, the central question in Section 3 is: no matter whether the initial and boundary conditions are in equilibrium (the detailed information does not need to be known), without extensive particle-particle interaction, can the steady state be intrinsically nonequilibrium?

Recent progress in quantum nonequilibrium statistical mechanics has shed light on energy non-quantization, quantum gravity, quantum information theory, etc. [e.g., 27-31]. Its relationship with the non-Boltzmann $p_k$ of the non-thermodynamic nonequilibrium model in this manuscript remains to be seen.

## 5. Extended Discussion

The second law of thermodynamics was originally established within the framework of classical mechanics [32], before the development of quantum mechanics. Interestingly, as discussed in Section 1.2, its counterexamples were also recently demonstrated in classical mechanics [1-3]. Section 3 implies that this coincidence may have a deep reason: the non-thermodynamic phenomenon ($\Delta_{12} \neq \Delta_{21}$) appears to favor unquantized energy and localized wave packets.

### 5.1 Local states and quantum chaos

The classical mechanical models in [1-3] are either isolated or isothermal. Without extensive particle-particle collisions in the locally nonchaotic barriers, the inconsistency with the second law of thermodynamics is fundamentally caused by the intrinsically nonequilibrium steady state, i.e., the non-Boltzmann distributions of particle number density and kinetic temperature. For the quantum models in Section 3, on the one hand, a bound state is always in equilibrium, as the wave function is nonlocal; on the other hand, wave-packet scattering depends on the local boundary conditions, allowing for the nonequilibrium distribution. The effects of the scattering state and the bound state may be viewed from the perspective of information. If a quantum particle is in a bound state, the Schrödinger equation does not result in any loss of information, and interaction of the particle with a thermal wall changes the global state. In a scattering state, especially when $\lambda$ is small, the particle-wall collision is critical to the local state, and the information of the internal state evolution may not be preserved.

The critical role of the nonlocal nature of the wave function reminds us of quantum chaos. It is well known that quantum systems are generally less likely to be chaotic than classical mechanical systems [33,34]. Quantum chaos often arises in excited states or scattering states [35], such as a Rydberg atom in a strong external force field [36] or quantum scattering [37,38]. It may be described by the random-matrix theory [39,40]. The similarity between Section 3 and quantum chaos seems reasonable: only when chaos may occur, can its opposite condition (nonchaoticity) be nontrivial and have substantial consequences.

## 5.2 Boltzmann's H-theorem and the principle of maximum entropy

Boltzmann's H-theorem cannot be directly applied to quantum mechanics. In the hypothesis of molecular chaos [5], before a particle-particle collision, the state is described by the two-body probability density ($f_{ab}$). After the collision, $f_{ab}$ is replaced by $f_a f_b$, where $f_a$ and $f_b$ are the one-body probability densities of the two particles, respectively. For two quantum particles, such a mathematical treatment is counterintuitive, because the particles would be entangled upon interaction. If the state is described by $f_a f_b$ before the particle-particle interaction and by $f_{ab}$ afterwards, entropy would decrease [5].

In Equations (4) and (62), the principle of maximum entropy is effectively taken as an "axiom". It represents the basic concept that the maximized probability should be assigned to the most probable state. Thus, the concern about the H-theorem is unnecessary. In this context, with no energetic penalty, entropy decrease can be spontaneously induced by Equations (7) and (67). The additional constraints are associated with the creation of information, i.e., possible knowledge about the locally nonchaotic system that is less random than its fully chaotic counterpart [1,2].

In Sections 2 and 3, the heat-engine statement of the second law of thermodynamics is analyzed through $\Delta_{12} = \Delta_{21}$ (Equation 14). As demonstrated in [1-3], Equation (14) is a powerful tool to examine whether the system is non-thermodynamic and whether the associated energy properties are nontrivial. Regardless of the detailed evolution process of the wave function, we monitor the overall system behavior. For a system/ensemble consisting of a large number of particles, the energetic penalty of the measurement of macroscopic variables (e.g., $F_v$ and $F_c$) is negligible. In equilibrium thermodynamics, Equations (4) and (14) are equivalent to each other, which is worth revisiting if the steady state is intrinsically nonequilibrium [1,2].

### 5.3 Considerations for future study

Clearly, the current case study of single-particle models does not provide a complete theoretical framework. Many questions are still unanswered. Extensive research needs to be carried out on quantum measurement, quantum chaos, quantum entanglement, nonchaotic environment, von Neumann entropy and density matrix, among others. For instance, in a quantum stadium [41], the quantum scars correspond to the classical periodic orbits. Can this correlation be related to $\bar{F}_j(p_k, \psi_k)$ (Equation 67)? At the fundamental level, what is the link between spontaneity and nonequilibrium [42]? What is the role of the Ehrenfest time? How do we characterize the thermodynamic properties of quantum entanglement and disentanglement, especially if no energy or information transfer is directly involved?

The non-thermodynamic mechanisms may have applications in a number of classical mechanical or "semiclassical" systems in energy science, condensed matter physics, etc. Priority should be given to the scattering states of localized wave packets, particularly when the particles or quasiparticles are non-interacting or weakly/sparsely interacting. For example, the upper limit of the power density of an intrinsically nonequilibrium free-electron Fermi gas could exceed 10 kW/cm$^3$ [1,2].

## 6. Concluding Remarks

Current research builds upon the recent finding of the fundamentals of statistical mechanics [1-3]: in classical mechanics, with locally nonchaotic components for which Boltzmann's H-theorem does not apply, certain large-sized models are intrinsically nonequilibrium. These systems may produce useful work by absorbing heat from a single thermal reservoir without any other effect, thereby breaking the boundaries of the second law of thermodynamics.

Here, the investigation is extended to quantum mechanics. The heat-engine statement of the second law of thermodynamics is formulated as the generalized Maxwell's relations, $\Delta_{12} = \Delta_{21}$ (Equation 14). When the model is in contact with a thermal reservoir, a bound state always satisfies Equation (14), while a scattering state may not. That is, compared to the classical mechanical systems, with quantized energy, the quantum systems are more likely to comply with the second law of thermodynamics. The non-thermodynamic behavior ($\Delta_{12} \neq \Delta_{21}$) favors unquantized energy and localized wave packets, exhibiting a tendency to be a "semiclassical" phenomenon.

The analysis of the Gibbs entropy (Equation 59) demonstrates that $\Delta_{12} \neq \Delta_{21}$ is in line with the maximization of entropy. In other words, it is the conventional statement of the second law of thermodynamics ($\Delta_{12} = \Delta_{21}$) that contradicts the principle of maximum entropy. With the additional constraints on $p_k$ (Equation 67), entropy is maximized to the local maximum in the phase space ($S_{\text{ne}}$), smaller than the global maximum at thermodynamic equilibrium ($S_{\text{eq}}$).

The inconsistency with the second law of thermodynamics arises from the lack of complete communication among the local parts of the system. Thermodynamic equilibrium refers to the global state. For instance, in a classical ideal gas, the steady-state particle number density of one region is correlated with that of another region by the Boltzmann factor, regardless of the distance between the two regions. In Equation (14), the points of application of the thermodynamic forces associated with $\Delta_{12}$ and $\Delta_{21}$ ($F_1$ and $F_2$) can be far apart. Interestingly, the reasons why classical mechanical systems and quantum systems obey $\Delta_{12} = \Delta_{21}$ appear to be different:

- In classical mechanics, according to the hypothesis of molecular chaos, the widespread particle-particle collisions are the root cause of the loss of local characteristics of individual particle movements.

- In quantum mechanics, the key factor is the innate nonlocal nature of the wave function, for the bound state (Section 3.2) and the scattering state with the thermal wavelength ($\lambda$) larger than the characteristic length ($L_0$) (Section 3.3).

- When the quantum particles are in thermal equilibrium, the second law of thermodynamics is followed. In Section 2, the equations of degeneracy pressure are derived from $\Delta_{12} = \Delta_{21}$.

- However, if $\lambda \ll L_0$ and thermal equilibrium cannot be reached, the quantum free particle in Section 3.3 can cause $\Delta_{12} \neq \Delta_{21}$, similarly to the classical mechanical models in Section 3.1.3 and [1-3]. The underlying reason is the dependence of the probability of quantum states ($p_k$) on the speed of wave packet (Equation 68). When the particle collides with a thermal wall, the internal state evolution is spontaneously and nonuniformly interrupted by the environment (a thermal reservoir).

## Appendix

### 1. Cross-influence of thermodynamic forces

Figure 4 depicts an isothermal operation cycle of a thermodynamic system. The system is closed and immersed in a thermal reservoir. There are two thermally correlated thermodynamic forces, $F_1$ and $F_2$. Their conjugate variables are $x_1$ and $x_2$, respectively.

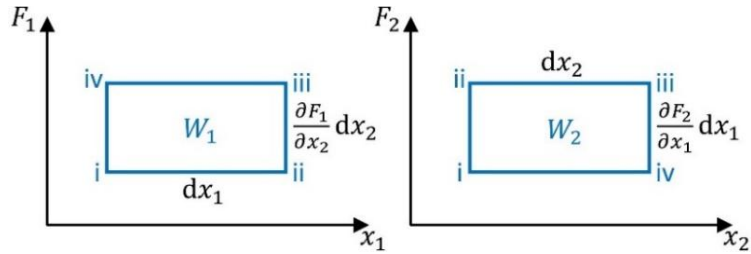

**Figure 4.** An isothermal operation cycle of two thermally correlated thermodynamic forces, $F_1$ and $F_2$. Indexes i-iv indicate the system states. Overall, $F_1$ produces work $W_1$, and $F_2$ consumes work $W_2$. The heat-engine statement of the second law of thermodynamics demands that $W_1 = W_2$, which leads to Equation (14).

From State i to ii, $x_1$ is increased by an arbitrarily small amount $dx_1$. It consumes work $F_1 dx_1$ and causes $F_2$ to change by $\Delta_{21} dx_1$ ($\Delta_{21} \triangleq \partial F_2 / \partial x_1$). From State ii to iii, $x_2$ is increased by an arbitrarily small amount $dx_2$. Correspondingly, the system consumes work $(F_2 + \Delta_{21} dx_1) dx_2$, and $F_1$ varies by $\Delta_{12} dx_2$ ($\Delta_{12} \triangleq \partial F_1 / \partial x_2$). From State iii to iv, $x_1$ decreases by $dx_1$; the system produces work $(F_1 + \Delta_{12} dx_2) dx_1$, and $F_2$ changes back. Finally, from State iv to i, $x_2$ decreases by $dx_2$; the system produces work $F_2 dx_2$ and returns to the initial state. In a complete cycle, $F_1$ produces work $W_1 = \Delta_{12} dx_2 dx_1$, and $F_2$ consumes work $W_2 = \Delta_{21} dx_1 dx_2$.

The operation in Figure 4 is reversible. The heat-engine statement of the second law of thermodynamics demands that no useful work can be produced in a cycle by absorbing heat from a single thermal reservoir without any other effect, i.e., $W_1 = W_2$. This requirement leads to $\Delta_{12} = \Delta_{21}$ (Equation 14).

For an equilibrium system, $\Delta_{12} = \Delta_{21}$ may be directly derived from $F_1 = \partial \mathcal{F} / \partial x_1$ and $F_2 = \partial \mathcal{F} / \partial x_2$, so that $\partial F_1 / \partial x_2 = \partial F_2 / \partial x_1 = \partial^2 \mathcal{F} / (\partial x_1 \partial x_2)$, with $\mathcal{F}$ being the Helmholtz free energy. Equation (14) can be viewed as the generalized Maxwell's relations.

## 2. Simplified form of the degeneracy pressure of Fermi gas

Equation (26) suggests that the degeneracy pressure ($P$) depends on the following three variables: $\beta^{1/2}/D_2$, $D_2/(\beta^{3/2}\rho)$, and $\rho$. Through dimensional analysis, we have $\bar{P} = \eta(x_F)$, where $\bar{P} = D_2 P/(\rho^2 \beta^{1/2})$, $\eta$ indicates a certain function of $x_F$, and $x_F = D_2/(\beta^{3/2}\rho)$. If $\eta(x_F) = x_F$, $\bar{P} = \eta(x_F)$ is the ideal gas law $P = \rho/\beta$; if $\eta(x_F) = \bar{\alpha}_1 \bar{x} + \bar{\alpha}_5 \bar{x}^5$, $\bar{P} = \eta(x_F)$ is the conventional approximate equation [10]: $P \approx [2N\epsilon_F/(5V_m)](1 + 5\pi^2 \bar{T}^2/12)$, where $\bar{x} = x_F^{1/3}$, $\bar{\alpha}_1 = 2s_k^{2/3}/5$, $\bar{\alpha}_5 = \pi^2 s_k^{-2/3}$, $\epsilon_F$ is the Fermi energy, $\bar{T} = T/T_F$, $T_F$ is the Fermi temperature, and $s_k = 3s_0/4$. Following this pattern, $\eta(x_F)$ may take the form of a Taylor series, so that $\bar{P} \approx \sum_{\xi=1}^{\infty} \mu_\xi \bar{x}^\xi$, with $\mu_\xi$ ($\xi = 1,2,3 ...$) being the coefficients.

The power of $\bar{x}$ cannot be larger than 5, to avoid ill-defined pressure at $\rho = 0$; that is, $\mu_\xi = 0$ for $\xi > 5$. When $\bar{x}$ is large ($\rho$ is small or $T$ is high), $P$ should converge to the ideal gas law ($\bar{P} = \alpha_3 \bar{x}^3$). Hence, $\mu_3 \approx 1$, $\mu_4 \approx 0$, and $\mu_5 \approx 0$. When $\bar{x}$ is small ($\rho$ is large or $T$ is low), $P$ should converge to the conventional approximate equation, so that $\mu_1 \approx \bar{\alpha}_1$. Consequently, $\bar{P} = \eta(x_F)$ becomes $\bar{P} \approx 2s_k^{2/3}\bar{x}/5 + \alpha_2 \bar{x}^2 + \bar{x}^3$, which can be rewritten as Equation (27), with $\mu_2 \approx -0.35$ being obtained from data fitting.

## 3. Two-dimensional quantum gas

Figure 5(a) shows a two-dimensional (2D) quantum gas. The system is closed and immersed in a thermal reservoir. In the interior, there is a high-potential zone. The potential difference between the high-potential zone and the outer zone is denoted by $u$. The total particle number ($N_{tot}$) and the high-potential zone area ($A_0$) are fixed. The potential difference between the two zones ($u$) and the outer-zone area ($A$) can be adjusted independently. The high-potential zone is an embodiment of the internal thermodynamic processes. It is used as an auxiliary element in the thermodynamic analysis. Neither $u$ nor $A_0$ will affect the equations of degeneracy pressure to be derived; both $u$ and $A_0$ can be arbitrarily small.

There are two thermally correlated thermodynamic forces. One is the degeneracy pressure in the outer zone ($P$), with the conjugate variable being $-A$. The other is the "support force" of the high-potential zone, $F_V$. It can be written as $F_V = F_0 N_V$, where $F_0$ is the contribution of a single particle, and $N_V$ is the particle number in the high-potential zone. To increase $u$ by a small amount ($du$), the system consumes work $F_V dz$, and vice versa, where $dz = du/F_0$, and $z = u/F_0$ is the conjugate variable of $F_V$. By setting $F_1 = P$ and $F_2 = F_V$ ($x_1 = -A$ and $x_2 = z$), Equation (14) becomes

$$-\frac{\partial F_V}{\partial A} = F_0 \frac{\partial P}{\partial u} \tag{69}$$

Figure 5(b) shows an operation cycle similar to Figures 2(b), 3(b), and 4. It indicates that Equation (69) represents the heat-engine statement of the second law of thermodynamics.

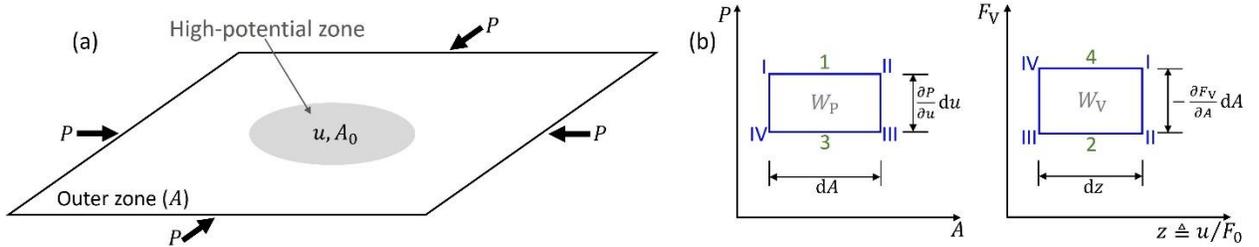

**Figure 5. (a)** Schematic of a 2D quantum gas. The system is closed and immersed in a thermal reservoir. In the interior, there is a high-potential zone (the grey circle). The total particle number ($N_{tot}$) and the high-potential zone area ($A_0$) are constant; the potential difference ($u$) and the outer-zone area ($A$) can be adjusted independently. The high-potential zone is used as an auxiliary element in the thermodynamic analysis; $u$ and $A_0$ can be arbitrarily small, and do not affect the derived equations of $P$. **(b)** An isothermal cycle, in which $A$ and $u$ are adjusted alternately. Indexes I-IV indicate the system states; numbers 1-4 indicate the operation steps; $dA$ and $dz$ are arbitrarily small. The heat-engine statement of the second law of thermodynamics dictates that the overall produced work ($W_P$) must be equal to the overall consumed work ($W_V$), which leads to Equation (69).

Denote the particle number in the outer zone by $N$. If temperature ($T$) and the particle number density ($\rho = N/A$) are known, $P$ is determined. In other words, $P$ is explicitly dependent on $\rho$ and $T$: $P = P(\rho, T)$; $u$ is an implicit variable, affecting $P$ through $\rho$. This relationship can be described by Equation (16), i.e.,

$$\frac{\partial P}{\partial \rho} = \frac{\frac{\partial P}{\partial u}}{\frac{\partial \rho}{\partial u}} \tag{70}$$

For a 2D Bose gas, the density of states is $g_i = mA/(2\pi\hbar^2)$ (for the outer zone) or $mA_0/(2\pi\hbar^2)$ (for the high-potential zone) [10], and the Bose-Einstein distribution is $f_{BE} = n_i/g_i = 1/(Be^{\beta\epsilon_i} - 1)$ (Equation 13). For the sake of simplicity, we assume that $T$ is sufficiently high and no Bose-Einstein condensate is formed. The particle numbers in the outer zone and in the high-potential zone are, respectively,

$$N = \int_0^\infty g_i f_{\text{BE}} \mathrm{d}\epsilon_i = C_{\text{B}} A \cdot \ln \frac{B}{B-1} \tag{71}$$

$$N_{\text{V}} = \int_u^\infty g_i f_{\text{BE}} \mathrm{d}\epsilon_i = C_{\text{B}} A_0 \cdot \ln \frac{B e^{\beta u}}{B e^{\beta u} - 1} \tag{72}$$

where $C_{\text{B}} = m/(2\beta\pi\hbar^2)$. Hence,

$$N_{\text{tot}} = N + N_{\text{V}} = C_{\text{B}} \left[ A \cdot \ln \frac{B}{B-1} + A_0 \cdot \ln \frac{B e^{\beta u}}{B e^{\beta u} - 1} \right] \tag{73}$$

Based on Equation (71),

$$B = \frac{\exp(C_{\text{B}}^{-1} \rho)}{\exp(C_{\text{B}}^{-1} \rho) - 1} \tag{74}$$

On both sides of Equation (73), taking the derivative with respect to $A$ leads to

$$\frac{\partial B}{\partial A} = B \left( \frac{A}{B-1} + \frac{A_0}{B e^{\beta u} - 1} \right)^{-1} \ln \frac{B}{B-1}, \tag{75}$$

and taking the derivative with respect to $u$ leads to

$$\frac{\partial B}{\partial u} = -\beta B \left[ \frac{A}{A_0} \frac{B e^{\beta u} - 1}{B - 1} + 1 \right]^{-1}. \tag{76}$$

Since $F_{\text{V}} = F_0 N_{\text{V}}$, with Equation (72), Equation (69) can be rewritten as

$$\frac{\partial P}{\partial u} = -\frac{\partial N_{\text{V}}}{\partial A} = \frac{C_{\text{B}} A_0}{B(B e^{\beta u} - 1)} \frac{\partial B}{\partial A} \tag{77}$$

Based on Equation (71),

$$\frac{\partial N}{\partial u} = -\frac{C_{\text{B}} A}{B(B-1)} \frac{\partial B}{\partial u} \tag{78}$$

Substitution of Equations (77) and (78) into Equation (70), combined with Equations (74-78), gives

$$\frac{\partial P}{\partial \rho} = \frac{1}{C_{\text{B}} \beta} \frac{\rho}{\exp(C_{\text{B}}^{-1} \rho) - 1} \tag{79}$$

Because $P \to 0$ when $\rho \to 0$, the degeneracy pressure is

$$P = \int_0^{N/A} \frac{\partial P}{\partial \rho} \mathrm{d}\rho = C_{\text{B}} k_{\text{B}} T \cdot P_{\text{n}}(x_{\text{u}}) \tag{80}$$

where $P_{\text{n}}(x_{\text{u}}) \triangleq \int_0^{x_{\text{u}}} [y(e^y - 1)^{-1}] \mathrm{d}y = \text{Li}_2(e^{x_{\text{u}}}) + x_{\text{u}} \ln(1 - e^{x_{\text{u}}}) - x_{\text{u}}^2/2 - \pi^2/6$, and $x_{\text{u}} = \rho/C_{\text{B}}$.

For a 2D Fermi gas, $g_i = s_0 mA/(2\pi\hbar^2)$ (in the outer zone) or $s_0 mA_{\text{V}}/(2\pi\hbar^2)$ (in the high-potential zone) [10], and the Fermi-Dirac distribution is $f_{\text{FD}} = 1/(B e^{\beta\epsilon_i} + 1)$ (Equation 13), where $s_0$ is the factor of spin. The particle numbers are

$$N = \int_0^\infty g_i f_{\text{FD}} \mathrm{d}\epsilon_i = L_{\text{B}} A \cdot \ln \frac{B+1}{B} \tag{81}$$

$$N_V = \int_u^\infty g_i f_{FD} \, d\epsilon_i = L_B A_0 \ln \frac{Be^{\beta u}+1}{Be^{\beta u}} \tag{82}$$

$$N_{tot} = N + N_V = L_B \left[ A \cdot \ln \frac{B+1}{B} + A_0 \ln \frac{Be^{\beta u}+1}{Be^{\beta u}} \right] \tag{83}$$

where $L_B = s_0 m / (2\beta\pi\hbar^2)$. Taking the derivative with respect to $A$ and $u$ on both sides of Equation (83) gives

$$\frac{\partial B}{\partial A} = B \left( \frac{A}{B+1} + \frac{A_0}{Be^{\beta u}+1} \right)^{-1} \ln \frac{B+1}{B} \tag{84}$$

$$\frac{\partial B}{\partial u} = -\beta B \left[ \frac{A}{A_0} \frac{Be^{\beta u}+1}{B+1} + 1 \right]^{-1} \tag{85}$$

Based on Equations (69) and (85),

$$\frac{\partial P}{\partial u} = -\frac{\partial N_V}{\partial A} = \frac{L_B A_0}{B(Be^{\beta u}+1)} \frac{\partial B}{\partial A} \tag{86}$$

Based on Equation (81),

$$\frac{\partial N}{\partial u} = -\frac{L_B A}{B(B+1)} \frac{\partial B}{\partial u} \tag{87}$$

$$B = \left[ \exp\left( \frac{\rho}{L_B} \right) - 1 \right]^{-1} \tag{88}$$

Substitution of Equations (84-88) into Equation (70) results in

$$\frac{\partial P}{\partial \rho} = \frac{\rho(B+1)}{L_B \beta} = \frac{1}{L_B \beta} \frac{\rho \cdot \exp(L_B^{-1}\rho)}{\exp(L_B^{-1}\rho)-1} \tag{89}$$

Since $P = 0$ when $\rho = 0$, based on Equation (89), we have

$$P = \int_0^{N/A} \frac{\partial P}{\partial \rho} \, d\rho = \frac{L_B}{\beta} P_m(x_v) \tag{90}$$

where $\quad P_m(x_v) \triangleq \int_0^{x_v} [ye^y(e^y - 1)^{-1}] dy = \mathrm{Li}_2(e^{x_v}) + x_v \ln(1 - e^{x_v}) - \pi^2/6 = P_n(x_v) + x_v^2/2$, and $x_v = L_B^{-1}\rho$. It can be verified that Equations (80) and (90) are equivalent to the conventional equation $P = U/A$, where the internal energy is $U = \int_0^\infty g_i f_{BE} \epsilon_i \, d\epsilon_i$ (for Bose gas) or $\int_0^\infty g_i f_{FD} \epsilon_i \, d\epsilon_i$ (for Fermi gas).

## 4. Inconsistency with the second law of thermodynamics

For the setup in Section 3.1.3, as the operation cycle in Figure 3(b) is performed on the ensemble in Figure 3(c), the overall produced work is the summation of the contributions of the individual boxes. The ensemble-average $\Delta_{vb}$ and $\Delta_{bv}$ can be calculated as

$$\Delta_{vb} = \frac{1}{1+J_x} \int_V^\infty \tilde{\Delta}_{vb} \cdot p_E(E) dE + \frac{J_x}{1+J_x} \int_V^\infty \tilde{\Delta}_{vb} \cdot p_E(E-V) dE \tag{91}$$

$$\Delta_{bv} = \frac{1}{1+J_x} \int_V^\infty \tilde{\Delta}_{bv} \cdot p_E(E) dE + \frac{J_x}{1+J_x} \int_V^\infty \tilde{\Delta}_{bv} \cdot p_E(E-V) dE \tag{92}$$

where $\tilde{\Delta}_{vb}$ and $\tilde{\Delta}_{bv}$ indicate $\Delta_{vb}$ and $\Delta_{bv}$ of a specific $E$, given by Equations (40) and (41), respectively; $p_E(E) = e^{-\beta E}/\sqrt{\pi E k_B T}$ is the 1D Maxwell-Boltzmann distribution of particle energy. In Equations (91) and (92), only $E > V$ needs to be considered; the first term represents the effect of the particles coming from the thermal wall on the lower shelf, and the second term is for the particles from the thermal wall on the upper shelf.

Figure 3(d) compares $-F_0 \Delta_{bv}$ and $\Delta_{vb}$ as functions of $\beta V$. It shows that $-F_0 \Delta_{bv} \neq \Delta_{vb}$. In fact, with the parameters under investigation, even the signs of $-F_0 \Delta_{bv}$ and $\Delta_{vb}$ are opposite.

## 5. Wave function of the particle in a box

According to [15], for the quantum particle in Figure 3(e), in the low-potential section ($0 \leq x \leq b$), the wave function is $\psi_I = A_1 \sin k_1 x$; in the high-potential section ($b < x \leq x_0$), the wave function is $\psi_{II} = A_a \sin k_2 x + A_b \cos k_2 x$, where $k_1 = \sqrt{2mE}/\hbar$, $k_2 = \sqrt{2m(E-V)}/\hbar$, and $A_1$, $A_a$, and $A_b$ are the coefficients. The boundary condition at the potential barrier ($x = b$) is $A_1 \sin k_1 b = A_a \sin k_2 b + A_b \cos k_2 b$ and $A_1 k_1 \cos k_1 b = A_a k_2 \cos k_2 b - A_b k_2 \sin k_2 b$. Based on $\int_0^b |\psi_I|^2 dx + \int_b^{x_0} |\psi_{II}|^2 dx = 1$ and the boundary condition, $A_1 = \alpha_A A_a$, $A_a = \sqrt{1/\alpha_C}$, and $A_b = \alpha_D A_a$, where $\alpha_A = (\sin k_2 b + \alpha_D \cos k_2 b)/\sin k_1 b$, $\alpha_C = \alpha_A^2 (2bk_1 - \sin 2bk_1)/(4k_1) + \alpha_E/(4k_2)$, $\alpha_D = (\alpha_k k_2 \cos k_2 b - \sin k_2 b)/(\alpha_k k_2 \sin k_2 b + \cos k_2 b)$, $\alpha_k = \tan(k_1 b)/k_1$, and $\alpha_E = 2(\alpha_D^2 + 1)k_2 a + (\alpha_D^2 - 1)(\sin 2k_2 x_0 - \sin 2k_2 b) + 4\alpha_D(\cos^2 k_2 b - \cos^2 k_2 x_0)$.

At $x = x_0$, the boundary condition is $A_a \sin k_2 x_0 + A_b \cos k_2 x_0 = 0$, which can be rewritten as $\sin k_2 x_0 + \alpha_D \cos k_2 x_0 = 0$. It gives the eigen values of $E$ (Equation 50): $E_n = V + (\hbar^2/2mx_0^2)(n\pi - \text{atan}\,\alpha_D)^2$, where $n = 1,2,3 \ldots$

## 6. Limit between the bound state and the scattering state

To analyze Equation (57), the limit between the bound state and the scattering state may be related to the gap between $E_n$ and $E_{n\pm1}$ ($\delta E_n$). In Figure 3(e), a small fluctuation or uncertainty in $V$ ($\delta V$) corresponds to the resolution of $E_n$ ($\Delta E_n$). Based on Equation (51), for $n \geq 6$, $\delta E_n \approx E_{n+1} - E_n \approx \pi^2 \hbar^2 (2n+1)/(2mx_0^2)$ and $\Delta E_n \approx (a/x_0)\delta V$. When $x_0$ (or $L_x$ in Figure 3f) is sufficiently large (e.g., $L_x \gg \pi^2 \hbar^2 (2n+1)/(2ma \cdot \delta V)$), $\delta E_n \ll \Delta E_n$ and the particle could be regarded as being in a scattering state. When $x_0$ (or $L_x$ in Figure 3f) is relatively small, as $\delta E_n > \Delta E_n$, the particle is in a bound state.

## 7. Wave function of the quantum free particle

For the quantum free particle in Figure 3(f) ($E > V$), according to [16], the solutions of the time-independent Schrödinger equation in the low-potential section ($x \leq 0$) and the high-potential section ($x > 0$) are, respectively, $\psi_I = A_c e^{ik_1 x} + A_d e^{-ik_1 x}$ and $\psi_{II} = A_2 e^{ik_2 x}$, where $A_c$ is the coefficient in $x \leq 0$, $A_d = 2A_c/(1 + k_{21})$, $A_2 = A_c(1 - k_{21})/(1 + k_{21})$, and $k_{21} = k_2/k_1$. Since $A_c$ is not normalizable, both $k_1$ and $k_2$ are continuous.

## 8. Analysis of the Gibbs entropy

There are three reasons why Section 4 is based on the Gibbs entropy rather than the von Neumann entropy or the density matrix. Firstly, the Gibbs entropy (Equation 59) is defined by $p_k$ and does not depend on $|\psi_k\rangle$. For the single-particle model in Figure 3(f), the Gibbs entropy is sufficient for understanding how the system may be intrinsically nonequilibrium. Regardless of $|\psi_k\rangle$, $p_k$ could be non-Boltzmannian (Equation 68).

Secondly, Equation (59) is comparable to the classical mechanical definition of entropy in [1-3] (Equation 2). Here, the key issue is to evaluate whether the steady state is in equilibrium or nonequilibrium, i.e., whether $p_k$ is proportional to the Boltzmann factor $e^{-\beta \epsilon_k}$.

Thirdly, using the Gibbs entropy is conservative and convenient, because Equation (59) circumvents the detailed information of $|\psi_k\rangle$. If the specifics of $|\psi_k\rangle$ are explicitly involved in the calculation, the following question must be answered: how to justify that we have not overlooked the possible work consumption in the process of obtaining such information? This situation is somewhat similar to Maxwell's demon [8]. The nature of quantum measurement has not been adequately understood, including its energy and information properties.

In the future, to explore complicated quantum systems, the von Neumann entropy may be employed. Nevertheless, the understanding of $p_k$ (e.g., Equations 64-68) will remain useful.

In Sections 2 and 3, like the Gibbs entropy, the heat-engine statement of the second law of thermodynamics (Equation 14) does not directly deal with the details of the wave function. It observes the time-average energy exchanges at the macroscopic/ensemble level, without relying on the evolution of system state on the microscopic scale.